\documentclass[aps,prl,twocolumn,superscriptaddress,longbibliography,
amsmath,amssymb,floatfix,10pt]{revtex4-2}

\usepackage{comment}
\usepackage{graphicx}
\usepackage{bm}
\usepackage{booktabs, float}
\usepackage{mathtools}
\usepackage{xcolor}
\usepackage[colorlinks,linkcolor=blue,anchorcolor=blue,
citecolor=blue,urlcolor=blue]{hyperref}
\usepackage{physics}
\usepackage{inputenc}

\newcommand{\Cov}{\operatorname{Cov}}

\begin{document}

\title{Single-Ensemble Multiparameter Squeezing with Qudits}

\author{Xiaoshui Lin}
\affiliation{Department of Physics, Washington University, St. Louis, Missouri 63130, USA}

\author{Chunlei Qu}
\affiliation{Department of Physics, Stevens Institute of Technology, Hoboken, New Jersey 07030, USA}
\affiliation{Center for Quantum Science and Engineering, Stevens Institute of Technology, Hoboken, New Jersey 07030, USA}

\author{Chong Zu}
\affiliation{Department of Physics, Washington University, St. Louis, Missouri 63130, USA}
\affiliation{Center for Quantum Leaps, Washington University, St. Louis, Missouri 63130, USA}

\author{Chuanwei Zhang}
\email{chuanwei.zhang@wustl.edu}
\affiliation{Department of Physics, Washington University, St. Louis, Missouri 63130, USA}
\affiliation{Center for Quantum Leaps, Washington University, St. Louis, Missouri 63130, USA}

\begin{abstract}
Conventional spin squeezing enhances a single sensing channel.
Here, we show how internal qudit levels enable simultaneous multiparameter squeezing within one ensemble.
In two-component magnetometry, a qutrit sensor provides two orthogonal and weakly compatible channels.
A collective twisting interaction squeezes both responses while preserving joint attainability of the ultimate sensitivity.
The sensing gain is quantified by using a matrix generalization of the Wineland sensitivity that retains both noise correlations and cross-channel response.
An interaction-based echo amplifies the signal to overcome noise from a fixed local joint readout, yielding a simulated $13~\mathrm{dB}$ gain over the product-state standard quantum limit for $N=128$ qutrits.
More generally, we use the single-site quantum Fisher information matrix to select reference states and channel quadratures for prescribed sensing tasks.
The tangent geometry permits at most $d-1$ independent, weakly compatible channels around a common pure reference state for a $d$-level sensor.
Our work provides a constructive task-to-protocol map for multiparameter squeezing in a single qudit ensemble.
\end{abstract}

\maketitle

\textit{Introduction}: Quantum-enhanced sensing can surpass the standard quantum limit (SQL) of uncorrelated particles \cite{Degen2017Quantum, Pezze2018Quantum, Demille2024Quantum, Montenegro2025Review}.
Extending this advantage from single-parameter estimation to simultaneous sensing of multiple field components remains a central challenge in quantum metrology. This capability is important for applications such as vector magnetic-field sensing with nitrogen-vacancy centers \cite{Kitazawa2017Vector, Weggler2020Determination, Lamba2024Vector, Wang2025Simultaneous} and atomic magnetometers \cite{Zhang2024Single-beam, Pradhan2016Three, Meng2023Machine, Wang2025Pulsed}. In many existing implementations, however, the spins remain effectively uncorrelated, and the sensitivity is therefore SQL-limited.

Existing beyond-SQL strategies for multiparameter sensing mainly follow two routes. The first uses specially structured single-ensemble states, most notably 2-anticoherent states, which can, in principle, achieve Heisenberg-limited sensitivity for up to three parameters \cite{Kolenderski2008Optimal, Baumgratz2016Quantum, Albarelli2019Evaluating, Hou2020Minimal, Sidhu2021Tight}.
However, these states are difficult to prepare and read out, and are often fragile against decoherence \cite{Pezze2017Optimal, Yang2025Overcoming}.
The second route distributes different parameters across multiple ensembles or modes. In this approach, spatial resolution is reduced, and the enhanced estimation of global parameters requires inter-ensemble entanglement \cite{Komar2014Quantum, Proctor2018Multiparameter, Gessner2020Multiparamete, Li2026Multiparameter, Zhang2026Distributed, Pezze2025Distributed, Mamaev2025Non-Gaussian}, which is challenging to generate and maintain.

Here, we exploit internal qudit levels to enable simultaneous multiparameter squeezing within a single ensemble \cite{Collins2002Bell, Luo2019Quantum, Wang2020Qudits}.
We first recast conventional spin squeezing as noise redistribution within one sensing channel, then study two-component magnetometry with qutrits.
A common reference state supports two orthogonal, weakly compatible channels, whose response observables are simultaneously squeezed by a two-axis-twisting-like (TAT-like) interaction.
We quantify the gain using a matrix generalization of the Wineland sensitivity that accounts for correlated noise and cross-channel response.
Numerical simulations show beyond-SQL sensitivity in both components, while the twisting dynamics preserves joint attainability of the quantum bound.

We then address how this gain can be accessed in a single measurement setting.
A fixed local positive-operator-valued measure (POVM) estimates both response mean values from the same record but introduces additional noise.
Then, an interaction-based echo amplifies the signal response, enabling a simulated $13~\mathrm{dB}$ gain over the product-state SQL for $N=128$ qutrits with this readout.
The End Matter develops the general $d$-level construction, using the single-site quantum Fisher information matrix (QFIM) to select reference states and channel quadratures, and presents a trapped-ion realization with power-law interactions.
It also establishes the bound of $d-1$ independent, weakly compatible channels around a common pure reference state.

\textit{Conventional squeezing as a task-selected noise redistribution}:
Squeezing enhances sensitivity by reducing the readout noise relative to the signal response.
To identify these ingredients, consider a parameter encoded on one sensor by $u_\theta=e^{-i\theta\lambda}$, where $\lambda$ is traceless and Hermitian.
For a pure state $\ket\phi$, the quantum Fisher information (QFI) is $f(\phi)=4(\Delta\lambda)_\phi^2 = 4 \bra{\phi} (\lambda - \bra{\phi} \lambda \ket{\phi})^2 \ket{\phi}$, and the quantum Cramér-Rao bound (QCRB) for locally unbiased estimation is $(\Delta\theta)^2\geq1/f(\phi)$ \cite{Braunstein1997Statistical, Kurdzialek2023Measurement, Albarelli2019Evaluating, Szczykulska2016Multi, Liu2020Quantum, Gessner2018Sensitivity, Sidhu2021TightBounds}.
Maximizing $f(\phi)$ selects an optimal reference state $\ket\varphi$ for $u_\theta$, which has the optimal sensitivity.

A small signal of $\theta$ displaces the optimal reference state in an orthogonal direction as
\begin{equation}
 u_\theta\ket\varphi
 \simeq 
 (1 - i \theta \langle \lambda\rangle_{\varphi})\ket\varphi-i\theta\ket\nu+O(\theta^2),
\end{equation}
where $\ket\nu=(\lambda-\langle\lambda\rangle_\varphi)\ket\varphi$ and $\langle\varphi|\nu\rangle=0$.
Writing $(\Delta\lambda)_\varphi = \bra{\phi} (\lambda - \bra{\phi} \lambda\ket{\phi})^2\ket{\phi}>0$, the normalized tangent $\ket{\tilde\nu}=\ket\nu/(\Delta\lambda)_\varphi$ identifies the direction of this distinguishable change.
Together, $\ket\varphi$ and $\ket{\tilde\nu}$ define an effective two-level channel with operators
\begin{align}
 X=\ket{\varphi}\bra{\tilde\nu} +&\ket{\tilde\nu}\bra{\varphi},\quad Y=-i\ket{\varphi}\bra{\tilde\nu} +i\ket{\tilde\nu}\bra{\varphi},\nonumber\\
 Z&=\ket{\varphi}\bra{\varphi} -\ket{\tilde\nu}\bra{\tilde\nu}.
 \label{eq-single-channel-quadratures}
\end{align}
These obey the Pauli algebra $[X,Y]=2iZ$.
Up to the overall phase, the first-order response of $u_\theta$ to $\ket{\varphi}$ is equivalent to $e^{-i\tilde\theta X}\ket\varphi$, with $\tilde\theta=(\Delta\lambda)_\varphi\theta$.
Thus $X$ generates the local signal, $Y$ reads it out through $\langle Y\rangle_\theta\simeq-2\tilde\theta$, and $Z$ measures the transduction rate slope.
This is the one-channel picture underlying Ramsey interferometry.

For conventional qubit magnetometry, $\lambda=s_z=\sigma_z/2$, and we may choose the optimal reference $\ket\varphi=(\ket{\downarrow} + \ket{\uparrow})/\sqrt2$.
Then, we have $(\Delta\lambda)_\varphi=1/2$ and $(X,Y,Z)=(\sigma_z, -\sigma_y,\sigma_x)$.
For an ensemble with $N$ qubits, we define $\mathcal X=\sum_{j=1}^N X^{(j)}$, $\mathcal Y=\sum_{j=1}^N Y^{(j)}$, and $\mathcal Z=\sum_{j=1}^N Z^{(j)}$.
The parameter is encoded by $U_\theta=e^{-i\theta\mathcal X/2}$.
For any probe state, the signal slope at $\theta=0$ is $G=i\langle[\mathcal X/2,\mathcal Y]\rangle=-\langle\mathcal Z\rangle$.
For $G\ne0$, error propagation gives
\begin{equation}
 (\Delta\theta)^2=(\Delta\mathcal Y)^2/|G|^2=(\Delta\mathcal Y)^2/|\langle\mathcal Z\rangle|^2.
\end{equation}
The product reference $\ket\varphi^{\otimes N}$ has $(\Delta\mathcal Y)^2=N$ and $\langle\mathcal Z\rangle=N$, giving the SQL $(\Delta\theta)_{\rm SQL}^2=1/N$.
Comparison with this reference yields the Wineland spin squeezing parameter \cite{Wineland1992Spinsqueezing, Wineland1994Spinsqueezing, Kitagawa1993Squeezed, Ma2011Quantum},
\begin{equation}
\xi_R^2=\frac{(\Delta\theta)^2}{(\Delta\theta)_{\rm SQL}^2} =\frac{N(\Delta\mathcal Y)^2}{|\langle\mathcal Z\rangle|^2},
 \label{eq-task-selected-wineland-parameter}
\end{equation}
with $\xi_R^2<1$ certifying sensitivity beyond the SQL.
Squeezing redistributes fluctuations between $\mathcal X$ and $\mathcal Y$. 
Metrological gain occurs when the reduction in readout noise outweighs any loss of polarization and, hence, of signal slope.
For a pure probe, the QFI $F_Q=(\Delta\mathcal X)^2$ sets the ultimate bound $(\Delta\theta)^2\geq1/F_Q$, which needs not be reached by the chosen readout of $\mathcal Y$.
Multiparameter sensing asks whether several independent, weakly compatible channels can share one local reference.

\begin{figure}[t]
 \centering
 \includegraphics[width=0.98\linewidth]{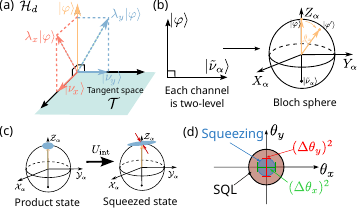}
 \caption{Task-selected sensing channels and multiparameter squeezing.
 (a) A generator $\lambda_\alpha$ displaces the reference state $\ket{\varphi}$
 along the orthogonal tangent $\ket{\nu_\alpha}$. For the qutrit prototype,
 $s_x$ and $s_y$ generate two orthogonal tangents from $\ket{0}$.
 (b) Each pair $\{\ket{\varphi},\ket{\tilde\nu_\alpha}\}$ defines a two-level channel with quadratures $X_\alpha,Y_\alpha,Z_\alpha$.
 The Bloch angle is $2\tilde\theta_\alpha$, with $\tilde\theta_\alpha=(\Delta\lambda_\alpha)_\varphi\theta_\alpha$.
 (c) Twisting $U_{\rm int}=e^{-iH_{\rm TAT}t}$ squeezes the responses $\mathcal Y_\alpha$ and anti-squeezes the conjugate signals $\mathcal X_\alpha$.
(d) Phase-estimation uncertainty ellipses for a multiparameter squeezed state and a direct-product state. The resulting joint sensitivity from squeezing exceeds that of the product-state SQL in all directions.}
 \label{Fig1-schematic}
\end{figure}

\textit{From one channel to two: qutrit magnetometry.}
To extend the one-channel construction, the two fields must produce independent local displacements, and their quantum-limited sensitivities must be jointly attainable.
We consider a two-component magnetometry with a qutrit, whose basis states $\ket+,\ket0,\ket-$ have $m_z=+1,0,-1$, with sensing operator, 
\begin{equation}
 u_{\boldsymbol\theta}=\exp(-i\theta_xs_x-i\theta_ys_y), \quad \boldsymbol\theta=(\theta_x,\theta_y).
 \label{eq-qutrit-encoding}
\end{equation}
We quote sensitivities per preparation at $\boldsymbol\theta_* =0$.
The estimation covariance $\Gamma_{\alpha\beta}=\Cov(\theta_\alpha,\theta_\beta)$ has diagonal entries $(\Delta\theta_\alpha)^2\equiv\Gamma_{\alpha\alpha}$ and off-diagonal entries describing correlated errors.
For a pure reference $\ket\phi$ with nonsingular QFIM, the covariance matrix obeys
\begin{equation}
 \Gamma\succeq f(\phi)^{-1},\qquad
 f_{\alpha\beta}(\phi)=4\Cov_\phi(s_\alpha,s_\beta),
 \label{eq-single-site-multiparameter-qcrb}
\end{equation}
where $\alpha,\beta=x,y$ and $A\succeq B$ means that $A-B$ is positive semidefinite.
In this manuscript, quantum covariances are symmetrized: $\Cov_\phi(A,B)=\langle\{A,B\}\rangle_\phi/2-\langle A\rangle_\phi\langle B\rangle_\phi$, with $\{A,B\}=AB+BA$.

Full rank of $f(\phi)$, with $\det[f(\phi)] \neq 0$, ensures that both parameters are locally identifiable.
For a pure state, the bound in Eq.~\eqref{eq-single-site-multiparameter-qcrb} is jointly attainable when weak commutativity holds \cite{Albarelli2019Evaluating, Szczykulska2016Multi, Liu2020Quantum, Gessner2018Sensitivity, Sidhu2021TightBounds, Matsumoto2002new, Pezze2017Optimal, Ragy2016Compatibility, Vidrighin2014Joint},
\begin{equation}
 i\langle[s_x,s_y]\rangle_\phi=-\langle s_z\rangle_\phi=0.
 \label{eq-qutrit-weak-commutativity}
\end{equation}
Thus, the compatibility depends on the reference state, even though the field generators do not commute.
A reference that improves one component need not improve the other.
For equal parameter weights, we therefore minimize the sum of the attainable error variances,
\begin{equation}
 C_q(\phi)=\Tr[f(\phi)^{-1}],
 \label{eq-qutrit-single-site-cost}
\end{equation}
over pure states with full-rank $f$ satisfying Eq.~\eqref{eq-qutrit-weak-commutativity}.
This selects $\ket\varphi=\ket0$ and $f_{\rm op}=f(\varphi)=4\mathbb I_2$ \cite{SI}.

The two channels now have a simple geometric meaning.
Acting on $\ket0$, the generators give two orthonormal tangents
\begin{align}
 \ket{\nu_x}&=(\ket++\ket-)/\sqrt2,\quad
 \ket{\nu_y}=i(\ket--\ket+)/\sqrt2,
 \label{eq-qutrit-tangents}
\end{align}
with $\langle\nu_\alpha|\nu_\beta\rangle=\delta_{\alpha\beta}$.
To first order, the encoded state is
\[
 u_{\boldsymbol\theta}\ket0
 \simeq\ket0-i\theta_x\ket{\nu_x}-i\theta_y\ket{\nu_y}.
\]
Each field component displaces the same reference along its own tangent [Fig.~\ref{Fig1-schematic}(a)].
The pairs $\{\ket0,\ket{\nu_x}\}$ and $\{\ket0,\ket{\nu_y}\}$ thus define two effective two-level channels that share one reference state.
The reference and two orthogonal tangents span all three qutrit levels here.
For a pure qubit, the analogous compatibility condition makes the inverse of the two-parameter QFIM singular \cite{SI}.
Thus, a qutrit is the minimal local sensor for this construction.

Applying Eq.~\eqref{eq-single-channel-quadratures} to each channel gives
\begin{align}
 X_x&=s_x, &Y_x&=-\{s_y,s_z\}, &Z_x&=s_y^2-s_z^2,\nonumber\\
 X_y&=s_y, &Y_y&=\{s_x,s_z\}, &Z_y&=s_x^2-s_z^2.
 \label{eq-qutrit-quadratures}
\end{align}
The responses $Y_x,Y_y$ are spin-quadrupole observables, quadratic in the spin components.
A small field creates coherences between $\ket0$ and the tangents, and $Y_x,Y_y$ detect their imaginary parts.
They play the same role as the transverse readout in conventional squeezing [Fig.~\ref{Fig1-schematic}(b)].
At the reference, $\langle Y_x\rangle_{\boldsymbol\theta}\simeq-2\theta_x$ and $\langle Y_y\rangle_{\boldsymbol\theta}\simeq-2\theta_y$.
Each response therefore measures one field component without cross-channel mixing.
In a general many-body state, both signal mixing and correlated noise must be retained.

For an ensemble of $N$ qutrits, we define $\mathcal X_\alpha=\sum_jX_\alpha^{(j)}$ and $\mathcal Y_\alpha=\sum_jY_\alpha^{(j)}$.
The encoding is $U_{\boldsymbol\theta}=\exp[-i(\theta_x\mathcal X_x+\theta_y\mathcal X_y)]$.
For a many-body state $\ket\Psi$, the two mean responses change as a vector,
\begin{equation}
\delta\langle\boldsymbol{\mathcal Y}\rangle\simeq G\boldsymbol\theta,\qquad
G_{\beta\alpha}=i\langle[\mathcal X_\alpha,\mathcal Y_\beta]\rangle_\Psi,
\end{equation}
where $\boldsymbol{\mathcal Y}=(\mathcal Y_x,\mathcal Y_y)^T$.
The diagonal entries of $G$ are the individual signal slopes; off-diagonal entries describe how one field affects the other response.
Similarly, $\mathcal C_{\alpha\beta}=\Cov_\Psi(\mathcal Y_\alpha,\mathcal Y_\beta)$ contains the response variances and their correlated noise.
For nonsingular $G$, multivariate error propagation gives the ideal collective-response benchmark \cite{Taylor2022Introduction, Young2025Engineering, Wu2025Spin}, which is denoted as generalized Wineland (GW) sensitivity in matrix form, as
\begin{equation}
 \Gamma_{\rm GW}=G^{-1}\mathcal C(G^T)^{-1}.
 \label{eq-qutrit-gw}
\end{equation}
The inverse slopes convert response noise into phase uncertainty, just as dividing by the squared slope does for a single parameter.
For the product reference $\ket{\Psi_{\rm p}}=\ket0^{\otimes N}$, $\mathcal C=N\mathbb I_2$ and $G=-2N\mathbb I_2$, giving the matrix version SQL $\Gamma_{\rm SQL}=\mathbb I_2/(4N)$.

The direct Wineland analogue compares the total phase uncertainty with this reference:
\begin{equation}
 \xi_{\rm GW}^2
 =\frac{\Tr\Gamma_{\rm GW}}{\Tr\Gamma_{\rm SQL}}
 =2N\Tr\Gamma_{\rm GW}.
 \label{eq-qutrit-normalized-gain}
\end{equation}
The condition $\xi_{\rm GW}^2<1$ certifies a reduction in the equally weighted sum of variances, but can conceal a tradeoff between directions.
Improvement in every parameter direction requires $\Gamma_{\rm SQL}\succ\Gamma_{\rm GW}$ or all eigenvalues of $4N\Gamma_{\rm GW}$ must be below unity.
We call this stronger condition direction-wise multiparameter squeezing.
As in the scalar case, the comparison accounts for both noise suppression and changes in signal slopes.
Weak compatibility does not imply that $\mathcal Y_x$ and $\mathcal Y_y$ commute, so realizing the ideal-response benchmark requires attention to the readout.
The local joint measurement discussed below introduces additional noise.

The state can contain more information than these responses capture.
For a pure probe, the many-body QFIM and its inverse are
\begin{equation}
 [F_Q]_{\alpha\beta}=4\Cov_\Psi(\mathcal X_\alpha,\mathcal X_\beta),
 \qquad \Gamma_Q=F_Q^{-1},
 \label{eq-qutrit-qfim}
\end{equation}
where $S_\alpha=\sum_j s_\alpha^{(j)}$ $(\alpha=x,y,z)$ are collective spins, and the channel signals coincide with the physical generators, $\mathcal X_x=S_x$ and $\mathcal X_y=S_y$.
The normalized intrinsic bound $\xi_Q^2=2N\Tr\Gamma_Q$ allows a direct comparison with $\xi_{\rm GW}^2$.
For nonsingular $F_Q$, $\Gamma_Q$ is jointly attainable when $i\langle[\mathcal X_x,\mathcal X_y]\rangle_\Psi=-\langle S_z\rangle_\Psi=0$.
The twisting dynamics below preserve this compatibility condition while generating squeezing.
The construction and concepts here can be generalized to broader multiparameter sensing tasks, as discussed in the End Matter.

\begin{figure}[h]
 \centering
 \includegraphics[width=0.99\linewidth]{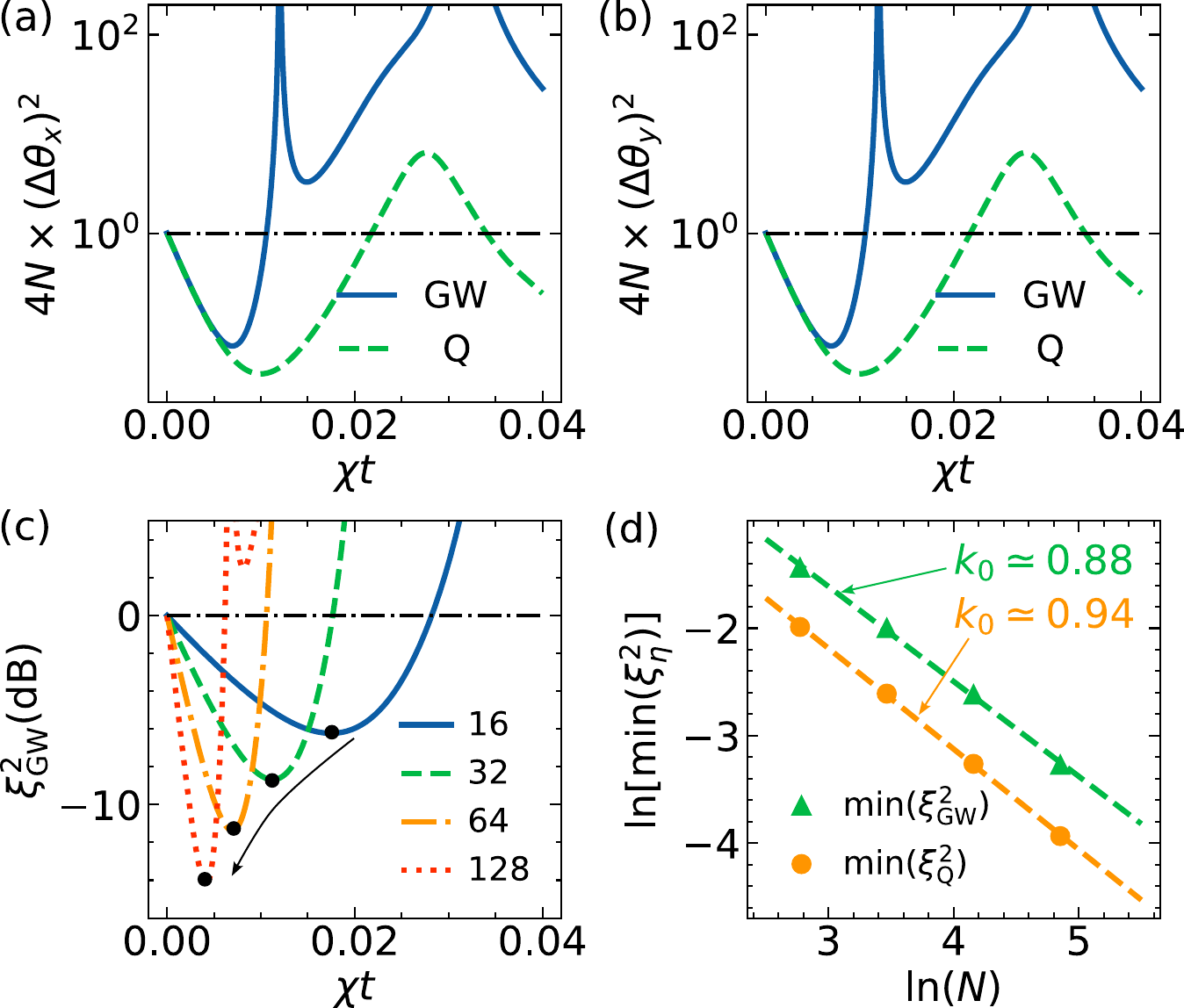}
 \caption{Simultaneous qutrit squeezing and metrological gain.
 (a)-(b) Sensitivities $4N(\Delta\theta_x)^2$ and
 $4N(\Delta\theta_y)^2$ versus $\chi t$ for $N=64$. GW denotes the collective-response sensitivity in Eq.~\eqref{eq-qutrit-gw}, Q denotes the intrinsic bound in Eq.~\eqref{eq-qutrit-qfim}, and the dash-dotted line is the product-state SQL.
 (c) $10\log_{10}\xi_{\rm GW}^2$ in dB for different $N$; dots mark the optimum for each size.
 (d) Size scaling of the optimized parameters.
 Fits to $\min(\xi_\eta^2)\simeq N^{-k_0}$ give $k_0\simeq0.88$ for $\eta={\rm GW}$ and $k_0\simeq0.94$ for $\eta=Q$.}
 \label{Fig2-Two-Parameter}
\end{figure}

\textit{Simultaneous squeezing generated by collective twisting:}
To squeeze both response quadratures, we evolve the product reference $\ket{\Psi_{\rm p}}$ with the TAT-like interaction
\begin{equation}
 H_{\rm TAT}=\chi\bigl(
 \{\mathcal X_x,\mathcal Y_x\}+\{\mathcal X_y,\mathcal Y_y\}\bigr).
 \label{eq-qutrit-tat}
\end{equation}
The collective products couple different qutrits, producing the probe $\ket{\Psi(t)}=e^{-iH_{\rm TAT}t}\ket{\Psi_{\rm p}}$.
In the large-$N$, short-time regime $0<N\chi t\ll1$, the channel algebra and truncated-Wigner analysis \cite{Polkovnikov2010Phase,Hosseinabadi2025User-Friendly,Schachenmayer2015Many-Body,Zhu201Generalized} in Ref.~\cite{SI} give
\[
 \mathcal C(t)\simeq N(1-8\chi Nt)\mathbb I_2,\qquad
 \xi_{\rm GW}^2(t)\simeq1-8\chi Nt.
\]
At this order, $G(t)\simeq-2N\mathbb I_2$.
The initial gain therefore comes from reducing both response noises while retaining their signal slopes; the conjugate signal fluctuations are amplified [Fig.~\ref{Fig1-schematic}(c) and (d)].
The interaction also conserves the total magnetization, $[H_{\rm TAT},S_z]=0$.
The initial value $\langle S_z\rangle_{\Psi_{\rm p}}=0$ is consequently preserved throughout the evolution.
Thus the compatibility condition established above continues to hold, and $\Gamma_Q$ remains jointly attainable whenever $F_Q$ is nonsingular.

Exact diagonalization extends this picture beyond the short-time approximation.
Both component variances fall below their product-state SQL in the same evolution window [Fig.~\ref{Fig2-Two-Parameter}(a) and (b)].
Optimizing the evolution time for each ensemble size gives finite-size fits $\min_t\xi_\eta^2\simeq N^{-k_0}$, with $k_0\simeq0.88\pm0.021$ for $\eta={\rm GW}$ and $k_0\simeq0.94\pm0.012$ for $\eta=Q$ [Fig.~\ref{Fig2-Two-Parameter}(c) and (d)].
Both the collective-response gain and the intrinsic quantum resource therefore improve with $N$ over the simulated range.
We next ask how much of this gain can be accessed with a fixed local joint measurement.

\textit{Joint response estimation and added noise from a local POVM:}
To estimate both response means in one setting, we apply the same single-site positive-operator-valued measure (POVM) to every qutrit.
The states $\ket{\varsigma_{\alpha,\pm}}=(\ket0\pm i\ket{\nu_\alpha})/\sqrt2$ are eigenstates of $Y_\alpha$ with eigenvalues $\pm1$.
A fixed six-outcome POVM is then \cite{Yuan2022Briefintroductionpovmmeasurement,Renes2004SIC-POVM,Geng2021What}
\begin{equation}
 \Pi_{\alpha,\pm}=\tfrac12\ket{\varsigma_{\alpha,\pm}}\!\bra{\varsigma_{\alpha,\pm}},
 \qquad \Pi_\pm=\tfrac12\ket\pm\!\bra\pm,
 \label{eq-endmatter-povm}
\end{equation}
where $\alpha=x,y$.
The two additional outcomes $\Pi_\pm$ complete the measurement, so that the six positive operators sum to $\mathbb I_3$.
For a local density matrix $\rho$, the probabilities $p_{\alpha,\pm}=\Tr(\rho\Pi_{\alpha,\pm})$ give
\begin{equation}
 \langle Y_\alpha\rangle=2(p_{\alpha,+}-p_{\alpha,-}).
 \label{eq-endmatter-povm-means}
\end{equation}
For each local outcome $m$, assign the score $\tilde y_\alpha(m)=\pm2$ to $(\alpha,\pm)$ and zero otherwise.
Summing over sites gives $\widehat{\mathcal Y}_\alpha=\sum_j\tilde y_\alpha(m_j)$, with $\mathbb E(\widehat{\mathcal Y}_\alpha)=\langle\mathcal Y_\alpha\rangle$.
Averaging these sums over repeated preparations therefore estimates both collective response means, even for entangled probes.
Ref.~\cite{SI} gives a Naimark implementation, where a fixed unitary in an enlarged $F=1\oplus F=2$ manifold maps the outcomes to distinguishable levels for Zeeman-resolved detection.

\begin{figure}[t]
 \centering
 \includegraphics[width=0.99\linewidth]{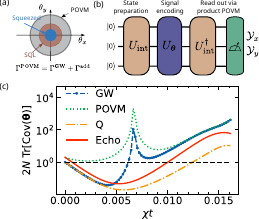}
 \caption{Fixed local POVM readout and interaction-based echo.
 (a) Phase-estimation ellipses for the ideal collective-response benchmark and the fixed local POVM; their covariance matrices differ by $\Gamma_{\rm add}$.
 (b) Twisting $U_{\rm int}$, encoding $U_{\boldsymbol\theta}$, reversal $U_{\rm int}^{\dagger}$, and local readout.
 (c) Normalized sensitivities $2N\Tr\Gamma_\eta$ for $\eta={\rm GW},{\rm POVM},Q,{\rm Echo}$ at $N=128$.
 The dashed line is the product-state SQL.}
 \label{Fig3-echo-readout}
\end{figure}

The mean estimates reproduce the slope matrix $G$, whereas their fluctuations depend on the second moments of the POVM scores.
Products of scores on distinct sites reproduce the corresponding quantum correlations, and the additional noise arises from the same-site terms \cite{SI}.
The classical covariance of the two estimates is therefore
\begin{equation}
 \mathcal C^{\rm POVM}=\mathcal C+\mathcal Q,\qquad
 \mathcal Q_{\alpha\beta}=\sum_j\langle\mathcal D_{\alpha\beta}^{(j)}\rangle,
 \label{eq-endmatter-output-covariance}
\end{equation}
where $\mathcal D_{\alpha\beta}=\sum_m\tilde y_\alpha(m)\tilde y_\beta(m)\Pi_m-\{Y_\alpha,Y_\beta\}/2$.
The added response noise satisfies $\mathcal Q\succeq0$.

Error propagation through nonsingular $G$ gives
\begin{equation}
 \Gamma_{\rm POVM}=\Gamma_{\rm GW}+\underbrace{G^{-1}\mathcal Q(G^T)^{-1}}_{\Gamma_{\rm add}}.
 \label{eq-povm-added-noise}
\end{equation}
The added phase covariance $\Gamma_{\rm add}$ enlarges the uncertainty ellipse [Fig.~\ref{Fig3-echo-readout}(a)] and can obscure the gain in $\Gamma_{\rm GW}$.
Even when $\xi_{\rm GW}^2<1$, the measured total phase variance may be above the product-state SQL.
However, it does not prevent the state-level metrological gain.
Instead, it indicates a more complex readout setup is needed to reveal the gain.

\textit{Signal amplification with an interaction-based echo:}
To address the issue of additional noise, we use an interaction-based echo \cite{Davis2019Approaching, Frowis2016Detecting, Nolan2017Optimal} to amplify the responses prior to the same local readout.
Starting from $\ket{\Psi_{\rm p}}$, we prepare the squeezed state, encode both phases, reverse the twisting evolution, and apply the POVM [Fig.~\ref{Fig3-echo-readout}(b)]:
\begin{equation}
 \ket{\Psi_{\rm echo}(\boldsymbol\theta)}
 =e^{iH_{\rm TAT}t}U_{\boldsymbol\theta}e^{-iH_{\rm TAT}t}\ket{\Psi_{\rm p}}.
 \label{eq-endmatter-echo-state}
\end{equation}
At $\boldsymbol\theta=0$, the forward and reverse evolutions cancel, returning the system to the product reference.
A small signal inserted between them produces a displacement governed by the transformed generators $\mathcal X_\alpha(t)=e^{iH_{\rm TAT}t}\mathcal X_\alpha e^{-iH_{\rm TAT}t}$, with slope matrix
\begin{equation}
 G^{\rm echo}_{\beta\alpha}
 =i\langle[\mathcal X_\alpha(t),\mathcal Y_\beta]\rangle_{\Psi_{\rm p}}.
 \label{eq-endmatter-echo-slope}
\end{equation}
Because the reversal is independent of the unknown phases, it preserves the QFIM of the encoded probe while changing its measured responses.

At the operating point, the product reference fixes both the intrinsic noise and the POVM contribution: $\mathcal C_{\rm p}=\mathcal Q_{\rm p}=N\mathbb I_2$.
The total output covariance is therefore $2N\mathbb I_2$, independent of the twisting time.
For nonsingular $G^{\rm echo}$, the phase sensitivity is
\begin{equation}
 \Gamma_{\rm echo}
 =(G^{\rm echo})^{-1}(2N\mathbb I_2)[(G^{\rm echo})^T]^{-1}.
 \label{eq-endmatter-echo-sensitivity}
\end{equation}
The echo improves sensitivity by increasing the signal slopes relative to this fixed output noise.
At $t=0$, the normalized cost $\xi_{\rm echo}^2=2N\Tr\Gamma_{\rm echo}=2$ reflects the readout penalty.
At suitable twisting times, the amplified slopes reduce this cost below unity.
For $N=128$, the optimized gain $-10\log_{10}\xi_{\rm echo}^2$ reaches $13~\mathrm{dB}$, compared with $14~\mathrm{dB}$ for GW and $17~\mathrm{dB}$ for QCRB [Fig.~\ref{Fig3-echo-readout}(c)].
The same fixed readout yields below-SQL sensitivity for all sizes simulated in Ref.~\cite{SI}.

\textit{Conclusion and discussion:}
Conventional spin squeezing redistributes noise within one signal-response channel.
A qudit supports several such channels around a common reference, allowing simultaneous multiparameter enhancement in a single ensemble.
For two-component magnetometry, the qutrit state $\ket0$ supplies two orthogonal, weakly compatible tangents; collective twisting squeezes both responses while preserving joint attainability, and an echo makes the gain accessible with a fixed local measurement.

The general construction in the End Matter extends this example to task-selected references, channels, and SQL benchmarks, with at most $d-1$ independent compatible channels.
Three-component examples for $d=4$ and $5$ are given in Ref.~\cite{SI}.
The trapped-ion example in the End Matter extends the preparation to power-law interactions. Further studies can address dimensionality, disorder, decoherence, and imperfect echo control \cite{Perlin2020Spin,Fossfeig2016entanglement,Block2024Scalable,Kaplanlipkin2025theory,Begg2026Scalablespin}, as well as other qudit platforms \cite{Wu2025Spinsqueezing,Barry2020Sensitivity,Mogerle2025Spin-1,VanBijnen2015Quantum,Qiao2025Realization,Champion2025Efficient,Neeley2009Emulation}.
The matrix-squeezing framework may also motivate witnesses for multipartite qudit entanglement \cite{Vitagliano2011SpinSqueezing,Vitagliano2014SpinSqueezing}.
\label{sec-main-end}

\begin{acknowledgments}
X. Lin thanks Samuel E. Begg and Jiahao Yan for insightful discussions.
X. Lin and C. Zhang acknowledge support from the National Science Foundation
under Grants No. OSI-2503230 and No. OSI-2228725. C. Qu is supported by
ACC-New Jersey under Contract No. W15QKN-18-D-0040. C. Zu acknowledges support
from the National Science Foundation under Grant No. 2514391.
\end{acknowledgments}

\bibliography{ref}

\clearpage
\appendix
\section{End Matter: general task-selected multiparameter squeezing}
\label{sec-general-theory}

\begin{table}[b]
\caption{\label{tab-correspondence}
Scalar-to-matrix structure of task-selected squeezing. The scalar response formula is recovered for $k=1$; $G$ always denotes the physical signal slope.}
\begin{ruledtabular}
\begin{tabular}{c@{\qquad}c@{\qquad}c}
 & $k=1$ & $k>1$ \\ 
 \hline
Noise & $(\Delta\mathcal Y)^2$ & $\mathcal C_{\alpha\beta} = \operatorname{Cov}(\mathcal{Y}_{\alpha},\mathcal{Y}_{\beta})$ \\
Slope & $G=i\langle[\Lambda,\mathcal Y]\rangle$ & $G_{\alpha\beta} =i\langle[\Lambda_\beta,\mathcal Y_\alpha]\rangle$ \\
Sensitivity & $(\Delta\mathcal Y)^2/|G|^2$ & $G^{-1}\mathcal C(G^T)^{-1}$ \\
Gain & $\Gamma_{\rm \eta}/\Gamma_{\rm SQL}$ & $\Tr(\Gamma_\eta)/\Tr(\Gamma_{\rm SQL})$
\end{tabular}
\end{ruledtabular}
\label{table-singel-multiparameter-comparison}
\end{table}

The qutrit prototype relies on three ingredients: independent tangent directions, weak compatibility, and a conjugate response for each signal.
These ingredients arise naturally for a prescribed $k$-parameter unitary task on one qudit $ u_{\boldsymbol\theta}=\exp\!\left(-i\sum_{\alpha=1}^{k}\theta_\alpha\lambda_\alpha\right)$.
For a pure reference state $\ket{\phi}$, we define the tangents orthogonal to the reference 
\begin{equation}
 \ket{\nu_\alpha(\phi)}= (\lambda_\alpha-\langle\lambda_\alpha\rangle_\phi)\ket{\phi},
 \label{eq-general-tangents}
\end{equation}
with their Gram matrix $ \Omega_{\alpha\beta}(\phi)= \langle\nu_\alpha(\phi)|\nu_\beta(\phi)\rangle.$
Its real and imaginary parts have distinct metrological roles,
\begin{align}
 f_{\alpha\beta}(\phi)
 &=4\operatorname{Re}\Omega_{\alpha\beta}(\phi)
 =4\Cov_\phi(\lambda_\alpha,\lambda_\beta),
 \nonumber\\
 \omega_{\alpha\beta}(\phi)
 &=-2\operatorname{Im}\Omega_{\alpha\beta}(\phi)
 =i\langle[\lambda_\alpha,\lambda_\beta]\rangle_\phi.
 \label{eq-general-geometry}
\end{align}
The metric $f$ is the single-site QFIM, whereas $\omega$ measures the weak compatibility of the parameter directions.
For a pure state, the QCRB is regular and jointly attainable when
\begin{equation}
 \det f(\phi)\neq 0,  \quad \omega_{\alpha\beta}(\phi)=0, \quad\forall~\alpha,\beta.
 \label{eq-general-compatibility}
\end{equation}
A pure $d$-level qudit has a tangent space of real dimension $2(d-1)$.
Weak compatibility makes the span of the tangent vectors an isotropic subspace, whose real dimension is at most $d-1$.
A full-rank QFIM further requires the $k$ tangent directions to be independent.
Consequently,
\begin{equation}
 k\le d-1
 \label{eq-channel-bound}
\end{equation}
is the maximum number of independent, weakly compatible local sensing channels.

Because the geometry depends on the reference state, for equal parameter weights $W=\mathbb I$ we select
\begin{align}
 \ket{\varphi}&\in \arg \{ \min_{\ket{\phi}} \Tr[f(\phi)^{-1}]\},
 \label{eq-general-reference-sql}
\end{align}
with $\Gamma_{\rm SQL}=f_{\rm op}^{-1}/N$, $C_{\rm SQL}=\Tr\Gamma_{\rm SQL}$, and $f_{\rm op}=f(\varphi)$, subject to Eq.~\eqref{eq-general-compatibility}.
Within a compatible subspace, $\Omega(\varphi)$ is real and positive definite.
The compatible tangents need not be mutually orthogonal. Each direction nevertheless defines the same channel operators $(X_\alpha, Y_\alpha, Z_\alpha)$ as in Eq.~\eqref{eq-single-channel-quadratures}.
The physical generators need not equal the normalized channel signals on the full Hilbert space.
With  $P_0=\ket\varphi\bra\varphi$, and $Q=\mathbb I-P_0$, their exact decomposition is
\begin{equation}
 \lambda_\alpha=(\Delta\lambda)_\varphi X_\alpha+R_\alpha,\qquad
 R_\alpha=P_0\lambda_\alpha P_0+Q\lambda_\alpha Q.
 \label{eq-general-generator-decomposition}
\end{equation}
Only $X_\alpha$ connects the reference to its tangent, while $R_\alpha$ can affect the physical response and QFIM after many-body evolution \cite{SI}.
We therefore retain the task-selected responses $Y_\alpha$ and use the full physical generators $\lambda_\alpha$ in the transduction matrix and squeezing Hamiltonians.
The selected single-site operators determine the corresponding collective operators $\Lambda_{\alpha} = \sum_j \lambda_{\alpha}^{(j)}$, $\mathcal{Y}_{\alpha} = \sum_j Y_{\alpha}^{(j)}$, and $\mathcal{G}_{\beta\alpha} = i\sum_j[\lambda_{\alpha}^{(j)}, Y_{\beta}^{(j)}]$.
For nonsingular matrices, the collective responses and the pure-state QFIM define the sensitivities 
\begin{equation}
 \Gamma_{\rm GW}=G^{-1}\mathcal C (G^T)^{-1},\quad  \Gamma_Q = F_Q^{-1},   
\end{equation}
where $\mathcal C_{\alpha\beta}=\Cov_\Psi(\mathcal Y_\alpha,\mathcal Y_\beta)$, $G_{\beta\alpha}=\langle\mathcal G_{\beta\alpha}\rangle_\Psi$, and $[F_Q]_{\alpha\beta}=4\Cov_\Psi(\Lambda_\alpha,\Lambda_\beta)$. Here $\Cov(A,B)=\langle\{A,B\}\rangle/2-\langle A\rangle\langle B\rangle$.
The squeezing capacity is defined by comparing those sensitivities to their optimal product-state value, from which we define
\begin{equation}
 \xi_\eta^2 =\frac{\Tr\Gamma_\eta}{\Tr\Gamma_{\rm SQL}} = \frac{N}{\Tr(f_{\rm op}^{-1})}\Tr\Gamma_\eta, \qquad \eta\in\{{\rm GW},Q\}.
 \label{eq-qudit-normalized-gain}
\end{equation}
For $k=1$, the GW expression becomes $4N\Delta_1^2(\Delta\mathcal Y)^2/|G|^2$.
When $G=-2\Delta_1\langle\mathcal Z\rangle$, as in conventional qubit magnetometry, it gives Eq.~\eqref{eq-task-selected-wineland-parameter}; otherwise the physical commutator must be retained.
Table~\ref{table-singel-multiparameter-comparison} summarizes the scalar-to-matrix correspondence.
Thus the same single-site geometry specifies both what must be squeezed and how the resulting gain is evaluated in parameter space.
Natural candidate interactions for simultaneous squeezing are $H_{\rm TAT}=\chi\sum_{\alpha=1}^{k}\{\Lambda_\alpha,\mathcal Y_\alpha\}$ and $H_{\rm OAT}=\chi\sum_{\alpha}\Lambda_\alpha^2$.
For $H_\eta$ with $\eta\in\{{\rm TAT},{\rm OAT}\}$, the condition $[H_\eta,i[\Lambda_\alpha,\Lambda_\beta]]=0$ and $\langle i[\Lambda_\alpha,\Lambda_\beta]\rangle_{\Psi_{\rm p}} = 0$ preserves weak compatibility. A useful and jointly attainable gain also requires nontrivial dynamics, a nonsingular $G$, and a full-rank $F_Q$, all of which must be verified for the chosen task.
The single-site geometry thus specifies reference selection, candidate squeezing interactions, and collective responses, with the implemented readout noise assessed separately as in the qutrit example.
Three-component constructions with $d=4$ and $5$ are presented in Ref.~\cite{SI}. A related $d=4$ single-site sensing geometry was discussed in Ref.~\cite{Isogawa2026Entanglement-Assisted}; here it supplies channels for squeezing across an ensemble of such sensors.

\clearpage
\section{End Matter: trapped-ion realization with power-law interactions}
\label{sec-trapped-ion}

\begin{figure}[th]
 \centering
 \includegraphics[width=0.99\linewidth]{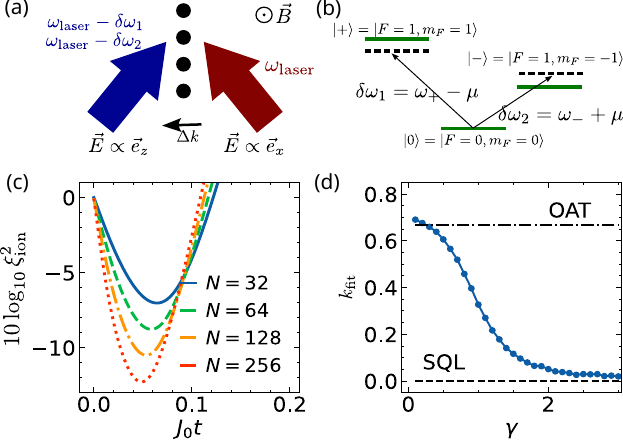}
 \caption{Trapped-ion implementation and finite-range squeezing.
 (a) Raman geometry for a linear ion chain.
 (b) Qutrit encoding and Raman beat notes; $\mu$ is the offset from the hyperfine transitions.
 (c) Optimized collective-response parameter $10\log_{10}\xi_{\rm ion}^2$ for $\gamma=0.5$, $V_0/J_0=2$, and $N=32,64,128,256$.
 (d) Finite-size exponent $k_{\rm fit}$ from $\min_t\xi_{\rm ion}^2\propto N^{-k_{\rm fit}}$ versus $\gamma$. The reference lines mark the qubit OAT exponent $2/3$ and zero scaling exponent.
 Numerical results use the discrete truncated Wigner approximation.}
 \label{Fig4-trapped-ion}
\end{figure}

The qutrit construction can also be explored with spatially decaying interactions in a trapped-ion chain, providing a route beyond the collective TAT model.
Consider $^{171}$Yb$^+$ ions in a linear Paul trap, with the effective spin-1 states encoded as $\ket\pm=\ket{F=1,m_F=\pm1}$ and $\ket0=\ket{F=0,m_F=0}$ in the $^2S_{1/2}$ manifold \cite{Senko2015Realization}.
A bias field resolves the two hyperfine transitions, whose frequencies are $\omega_\pm$.
Global Raman beams with transverse wave-vector difference $\Delta k$ provide beat notes $\omega_+-\mu$ and $\omega_-+\mu$ [Fig.~\ref{Fig4-trapped-ion}(a),(b)].
Off-resonant coupling to transverse phonons mediates approximately power-law spin exchange, with a tunable exponent $0\lesssim\gamma\lesssim3$.
For uniform effective light shifts, the ideal spin model is
\begin{equation}
 H_{\rm XY}=\sum_{i<j}\frac{J_0}{|i-j|^\gamma}
 \bigl(s_x^{(i)}s_x^{(j)}+s_y^{(i)}s_y^{(j)}\bigr)
 -V_0\sum_j(s_z^{(j)})^2.
 \label{eq-ion-xy}
\end{equation}
Here $V_0$ is the quadratic level shift; a uniform linear shift proportional to $S_z$ is inactive in the $S_z=0$ sector.
Residual inhomogeneous shifts and spin-phonon coupling are neglected in this model.

Since $[H_{\rm XY},S_z]=0$, evolution from $\ket{\Psi_{\rm p}}=\ket0^{\otimes N}$ remains in this sector and preserves weak compatibility for the two transverse signals.
At $\gamma=0$, the Hamiltonian becomes, up to a constant,
\begin{equation}
 H_{\rm XY}=\frac{J_0}{2}(S_x^2+S_y^2)
 +\left(\frac{J_0}{2}-V_0\right)\sum_j(s_z^{(j)})^2.
 \label{eq-ion-collective-limit}
\end{equation}
The collective term supplies OAT-like dynamics in both channels; the quadratic shift is retained in the simulations.
For $\gamma>0$, the nonuniform couplings break permutation symmetry, and the dynamics need not remain in the fully symmetric collective subspace.

\textit{Aligning the two squeezed responses.}
OAT-like evolution generally squeezes rotated quadratures.
A common phase rotation $R(\kappa)=\exp[-i\kappa\sum_j\ket0_j\!\bra0]$ aligns both channels before signal encoding:
\begin{equation}
 R^\dagger(\kappa)\mathcal Y_\alpha R(\kappa)
 =\cos\kappa\,\mathcal Y_\alpha+\sin\kappa\,\mathcal X_\alpha,
 \quad\alpha=x,y.
 \label{eq-ion-rotation}
\end{equation}

For the probe $\ket{\Psi_\kappa(t)}=R(\kappa)e^{-iH_{\rm XY}t}\ket{\Psi_{\rm p}}$, we evaluate the same physical slopes and response covariance as in the main text,
\begin{align}
 G_{\beta\alpha}(\kappa,t)
 &=i\langle[\mathcal X_\alpha,\mathcal Y_\beta]\rangle_{\Psi_\kappa(t)},\nonumber\\
 \mathcal C_{\alpha\beta}(\kappa,t)
 &=\Cov_{\Psi_\kappa(t)}(\mathcal Y_\alpha,\mathcal Y_\beta),\nonumber\\
 \xi_{\rm ion}^2(t)
 &=\min_\kappa 2N\Tr\!\left[G^{-1}\mathcal C(G^T)^{-1}\right].
 \label{eq-ion-squeezing}
\end{align}
The rotation commutes with $S_z$ and preserves compatibility.
Rotational symmetry makes the two transverse sensitivities equal, so $\xi_{\rm ion}^2<1$ improves both components relative to $\Gamma_{\rm SQL}=\mathbb I_2/(4N)$.

The discrete truncated Wigner simulations \cite{Schachenmayer2015Many-Body,Zhu201Generalized} show simultaneous squeezing for $\gamma=0.5$ and $V_0/J_0=2$ [Fig.~\ref{Fig4-trapped-ion}(c)], reaching about $12~\mathrm{dB}$ for $N=256$.
Fits of the minimum versus $N$ give $k_{\rm fit}\simeq0.69$ at $\gamma=0$; the exponent decreases toward zero as $\gamma$ approaches $3$ [Fig.~\ref{Fig4-trapped-ion}(d)].
These finite-size, semiclassical results support metrological gain with spatially decaying interactions.

The same readout strategy as in the main text applies after preparation with $U_{\rm prep}=R(\kappa)e^{-iH_{\rm XY}t}$.
The sequence is $U_{\rm rd}U_{\boldsymbol\theta}U_{\rm prep}\ket{\Psi_{\rm p}}$, followed by the fixed local POVM.
An exact echo uses $U_{\rm rd}=U_{\rm prep}^\dagger$; more generally, any parameter-independent readout unitary preserves the encoded QFIM.
To recover the collective-response gain, the readout should amplify both response quadratures before detection \cite{Davis2019Approaching,Nolan2017Optimal}.

\end{document}


\title{Supplemental Material for ``Single-Ensemble Multiparameter Squeezing with Qudits''}
\author{Xiaoshui Lin}
\affiliation{Department of Physics, Washington University, St. Louis, Missouri 63130, USA}

\author{Chunlei Qu}
\affiliation{Department of Physics, Stevens Institute of Technology, Hoboken, New Jersey 07030, USA}
\affiliation{Center for Quantum Science and Engineering, Stevens Institute of Technology, Hoboken, New Jersey 07030, USA}

\author{Chong Zu}
\affiliation{Department of Physics, Washington University, St. Louis, Missouri 63130, USA}
\affiliation{Center for Quantum Leaps, Washington University, St. Louis, Missouri 63130, USA}

\author{Chuanwei Zhang}
\email{chuanwei.zhang@wustl.edu}
\affiliation{Department of Physics, Washington University, St. Louis, Missouri 63130, USA}
\affiliation{Center for Quantum Leaps, Washington University, St. Louis, Missouri 63130, USA}

\date{\today}

\maketitle 

\tableofcontents

\clearpage

This Supplemental Material provides derivations and additional examples that support the main text and End Matter.
Secs. S1-S3 develop the operator conventions, tangent construction, and reference-state optimization.
Secs. S4-S5 analyze the qutrit twisting dynamics, and Secs. S6-S7 give the implementation, noise analysis, and finite-size results for the joint readout.
Sec. S8 connects the trapped-ion preparation in the last End Matter section to interaction-based readout, including the conditions under which the collective-response gain can be recovered.

We use the same notation as the main text.
$\ket\phi$ is a trial single-site state, $\ket\varphi$ is a selected optimal reference, and $\ket{\Psi_{\rm p}}=\ket\varphi^{\otimes N}$ is the product probe. 
Lowercase $u_{\boldsymbol\theta}$ denotes single-site encoding and uppercase $U_{\boldsymbol\theta}$ its collective counterpart. Local spins $s_\alpha$ and channel quadratures $X_\alpha,Y_\alpha,Z_\alpha$ give collective operators $S_\alpha$ and $\mathcal X_\alpha,\mathcal Y_\alpha,\mathcal Z_\alpha$ by summing over sites. The slope $G_{\beta\alpha}$ has response index $\beta$ and parameter index $\alpha$. We distinguish response covariance $\mathcal C$, added response noise $\mathcal Q$, and phase-estimation covariance $\Gamma_{\alpha\beta}=\Cov(\theta_\alpha,\theta_\beta)$; sensitivity labels appear as subscripts, as in $\Gamma_{\rm GW}$ and $\Gamma_{\rm echo}$. Quantum covariances are symmetrized, $\Cov(A,B)=\langle\{A,B\}\rangle/2-\langle A\rangle\langle B\rangle$, whereas covariances of the measured scores are classical. All local sensitivities are evaluated at $\boldsymbol\theta=0$ per preparation. We set $\hbar=1$ except where explicitly restored in the control Hamiltonians.


\section{Definition of the generalized Gell-Mann matrices and spin operators in a qudit}

We denote the symmetric, antisymmetric, and diagonal generalized Gell-Mann generators by $T^{(S)}_{\mu\nu}$, $T^{(A)}_{\mu\nu}$, and $H_\mu$, respectively:
\begin{align}
 T^{(S)}_{\mu\nu}&=|\mu\rangle\langle\nu|+|\nu\rangle\langle\mu|,
 &1\leq\mu<\nu\leq d,\nonumber\cr
 T^{(A)}_{\mu\nu}&=-i(|\mu\rangle\langle\nu|-|\nu\rangle\langle\mu|),
 &1\leq\mu<\nu\leq d,\nonumber\cr
 H_\mu&=\sqrt{\frac{2}{\mu(\mu+1)}}\left[\sum_{r=1}^{\mu}|r\rangle\langle r|-\mu|\mu+1\rangle\langle\mu+1|\right],
 &1\leq\mu\leq d-1.
 \label{eq-si-ggm-matrice}
\end{align}
Together, these $d^2-1$ traceless Hermitian matrices form an orthogonal operator basis, $\Tr(T_aT_b)=2\delta_{ab}$. For $d=2$, $T^{(S)}_{12}=\sigma_x$, $T^{(A)}_{12}=\sigma_y$, and $H_1=\sigma_z$. In the spin-$s$ representation,
\begin{equation}
 s_x=\frac12\sum_{\nu=1}^{d-1}\sqrt{(d-\nu)\nu}\,T^{(S)}_{\nu,\nu+1},\quad
 s_y=\frac12\sum_{\nu=1}^{d-1}\sqrt{(d-\nu)\nu}\,T^{(A)}_{\nu,\nu+1},\quad
 s_z=\sum_{\mu=1}^{d-1}c_\mu H_\mu,
 \label{eq-si-vector-elements}
\end{equation}
where $c_\mu=\Tr(s_zH_\mu)/2$. Consequently, $\Tr(s_\alpha s_\beta)=d(d^2-1)\delta_{\alpha\beta}/12$.

\section{Tangent-space decomposition and the many-body QFIM}
\label{sec-si-tangent-decomposition}

This section makes explicit the relation between the physical generators $\lambda_\alpha$ and the channel quadratures constructed in the End Matter. Fix the selected reference $|\varphi\rangle$ and define
\begin{equation}
 P_0=|\varphi\rangle\langle\varphi|,\qquad Q=\mathbb I-P_0,\qquad
 |\nu_\alpha\rangle=Q\lambda_\alpha|\varphi\rangle
 =(\lambda_\alpha-\langle\lambda_\alpha\rangle_\varphi)|\varphi\rangle.
 \label{eq-si-nu-alpha-def}
\end{equation}
For every nonzero tangent vector,
\begin{equation}
 \Delta_\alpha=(\Delta\lambda_\alpha)_\varphi=\sqrt{\langle\nu_\alpha|\nu_\alpha\rangle},\qquad
 |\tilde\nu_\alpha\rangle=|\nu_\alpha\rangle/\Delta_\alpha,
\end{equation}
and
\begin{equation}
 X_\alpha=|\varphi\rangle\langle\tilde\nu_\alpha|+|\tilde\nu_\alpha\rangle\langle\varphi|,\quad
 Y_\alpha=-i|\varphi\rangle\langle\tilde\nu_\alpha|+i|\tilde\nu_\alpha\rangle\langle\varphi|,\quad
 Z_\alpha=|\varphi\rangle\langle\varphi|-|\tilde\nu_\alpha\rangle\langle\tilde\nu_\alpha|.
 \label{eq-si-channel-quadratures}
\end{equation}
These operators obey the Pauli algebra within $\operatorname{span}\{|\varphi\rangle,|\tilde\nu_\alpha\rangle\}$. With
\begin{equation}
 R_\alpha=P_0\lambda_\alpha P_0+Q\lambda_\alpha Q,
\end{equation}
the exact generator decomposition is
\begin{equation}
 \lambda_\alpha=\Delta_\alpha X_\alpha+R_\alpha.
 \label{eq-si-lambda-as-x}
\end{equation}
Only the $X_\alpha$ block connects $|\varphi\rangle$ to its tangent space; $R_\alpha$ does not contribute to the first-order displacement or the single-site QFIM at $|\varphi\rangle$.

The tangent Gram matrix is
\begin{equation}
 \Omega_{\alpha\beta}=\langle\nu_\alpha|\nu_\beta\rangle
 =\Delta_\alpha\Delta_\beta
 \langle\tilde\nu_\alpha|\tilde\nu_\beta\rangle.
\end{equation}
Its real and imaginary parts reproduce the geometry defined in the End Matter,
\begin{equation}
 f_{\alpha\beta}=4\operatorname{Re}\Omega_{\alpha\beta},\qquad
 \omega_{\alpha\beta}=-2\operatorname{Im}\Omega_{\alpha\beta}
 =i\langle[\lambda_\alpha,\lambda_\beta]\rangle_\varphi.
 \label{eq-si-qfim-omega}
\end{equation}
Equivalently,
\begin{equation}
 f_{\alpha\beta}=4\Delta_\alpha\Delta_\beta
 \Cov_\varphi(X_\alpha,X_\beta).
\end{equation}
Thus, weak commutativity is $\omega_{\alpha\beta}=0$. As detailed in Sec.~S3, the reference $|\varphi\rangle$ is selected by minimizing $C_q(\phi)=\Tr[f(\phi)^{-1}]$ over trial states $|\phi\rangle$, subject to weak commutativity and full rank. The quadratures and tangent norms above are then fixed at $|\varphi\rangle$. In the scalar case, the same norm is denoted by $\Delta_\lambda$, as in the main text.

For an ensemble, define
\begin{equation}
 \Lambda_\alpha=\sum_j\lambda_\alpha^{(j)},\quad
 \mathcal X_\alpha=\sum_jX_\alpha^{(j)},\quad
 \mathcal Y_\alpha=\sum_jY_\alpha^{(j)},\quad
 \mathcal Z_\alpha=\sum_jZ_\alpha^{(j)},\quad
 \mathcal R_\alpha=\sum_jR_\alpha^{(j)}.
 \label{eq-si-collective-operators}
\end{equation}
The physical encoding is $U_{\boldsymbol\theta}=\exp[-i\sum_\alpha\theta_\alpha\Lambda_\alpha]$, and for a pure many-body probe $|\Psi\rangle$,
\begin{equation}
 [F_Q]_{\alpha\beta}=4\Cov_\Psi(\Lambda_\alpha,\Lambda_\beta).
\end{equation}
Using $\Lambda_\alpha=\Delta_\alpha\mathcal X_\alpha+\mathcal R_\alpha$ gives the exact decomposition
\begin{equation}
 \begin{aligned}
 [F_Q]_{\alpha\beta}={}&4\Delta_\alpha\Delta_\beta\Cov_\Psi(\mathcal X_\alpha,\mathcal X_\beta)
 +4\Delta_\alpha\Cov_\Psi(\mathcal X_\alpha,\mathcal R_\beta)\\
 &+4\Delta_\beta\Cov_\Psi(\mathcal R_\alpha,\mathcal X_\beta)
 +4\Cov_\Psi(\mathcal R_\alpha,\mathcal R_\beta).
 \end{aligned}
 \label{eq-si-many-body-qfim-decomp}
\end{equation}
Therefore, signal-quadrature anti-squeezing alone need not equal the QFIM of the physical generators. Because the cross-covariance terms in Eq.~\eqref{eq-si-many-body-qfim-decomp} need not be positive semidefinite, no general matrix ordering follows from this decomposition. In the qutrit example discussed in the main text, however, $R_\alpha=0$ and $\Delta_\alpha=1$ for $\alpha=x,y$, so
\begin{equation}
 [F_Q]_{\alpha\beta}=4\Cov_\Psi(\mathcal X_\alpha,\mathcal X_\beta).
\end{equation}
With $f_{\rm op}=f(\varphi)$, the common product-state benchmark is $\Gamma_{\rm SQL}=f_{\rm op}^{-1}/N$ and $C_{\rm SQL}=\Tr\Gamma_{\rm SQL}$. The intrinsic sensitivity and its SQL-normalized cost are
\begin{equation}
 \Gamma_Q=F_Q^{-1},\qquad
 \xi_Q^2=\frac{\Tr\Gamma_Q}{C_{\rm SQL}}
 =\frac{N\Tr\Gamma_Q}{\Tr(f_{\rm op}^{-1})}.
\end{equation}
This distinguishes the intrinsic resource $\xi_Q^2$ from the collective-readout gain $\xi_{\rm GW}^2$.
Both quantities use the same product-state benchmark, as in the End Matter.

\section{Optimization of the single-site QFIM for multiparameter sensing}

We consider the $k$-parameter unitary task
\begin{equation}
 u_{\boldsymbol\theta}=\exp\!\left(-i\sum_{\alpha=1}^{k}\theta_\alpha\lambda_\alpha\right)
\end{equation}
on a normalized qudit probe $|\phi\rangle=\sum_{\mu=1}^{d}z_\mu|\mu\rangle$. The tangents orthogonal to the reference and their Gram matrix are
\begin{equation}
 |\nu_\alpha(\phi)\rangle=(\lambda_\alpha-\langle\lambda_\alpha\rangle_\phi)|\phi\rangle,
 \qquad \Omega_{\alpha\beta}(\phi)=\langle\nu_\alpha(\phi)|\nu_\beta(\phi)\rangle.
\end{equation}
As in the End Matter, the single-site QFIM and compatibility form are
\begin{equation}
 f_{\alpha\beta}(\phi)=4\operatorname{Re}\Omega_{\alpha\beta}(\phi) =4\Cov_\phi(\lambda_\alpha,\lambda_\beta),
 \qquad
 \omega_{\alpha\beta}(\phi)=-2\operatorname{Im}\Omega_{\alpha\beta}(\phi) =i\langle[\lambda_\alpha,\lambda_\beta]\rangle_\phi.
\end{equation}
For equal parameter weights, the optimal reference state is
\begin{equation}
 C_q(\phi)=\Tr[f(\phi)^{-1}],\qquad
 |\varphi\rangle\in\underset{|\phi\rangle}{\arg\min}\,C_q(\phi),
\end{equation}
subject to $\det f(\phi) \neq 0$ and $\omega_{\alpha\beta}(\phi)=0$ for every pair. The corresponding product-state probe and task-specific SQL are
\begin{equation}
 |\Psi_{\rm p}\rangle=|\varphi\rangle^{\otimes N},\qquad
 \Gamma_{\rm SQL}=\frac{f_{\rm op}^{-1}}{N},\qquad
 C_{\rm SQL}=\Tr\Gamma_{\rm SQL},\qquad f_{\rm op}=f(\varphi).
\end{equation}
The following examples evaluate this optimization for one-, two-, and three-component magnetic-field sensing. The physical generators and their phase normalization are specified in each case.

\subsection{Scalar magnetic-field sensing}

For scalar sensing, we use the same generator as in the main text, $\lambda=s_z=\sigma_z/2$. Writing
$|\phi\rangle=\sin(\theta_0/2)e^{i\phi_0}|\downarrow\rangle+\cos(\theta_0/2)|\uparrow\rangle$, with $s_z|\uparrow\rangle=\tfrac12|\uparrow\rangle$ and $s_z|\downarrow\rangle=-\tfrac12|\downarrow\rangle$, gives
\begin{equation}
 f(\phi)=4(\Delta s_z)_\phi^2=\sin^2\theta_0.
\end{equation}
The maximum $f_{\rm op}=1$ is reached on the equator; choosing $\phi_0=0$ gives
\begin{equation}
 |\varphi\rangle=\frac{|\downarrow\rangle+|\uparrow\rangle}{\sqrt2},\qquad
 |\nu\rangle=\frac{|\uparrow\rangle-|\downarrow\rangle}{2\sqrt2},\qquad
 \Delta_\lambda=\frac12.
\end{equation}
Consequently,
\begin{equation}
 (X,Y,Z)=(\sigma_z,-\sigma_y,\sigma_x),\qquad
 \tilde\theta=\Delta_\lambda\theta=\frac{\theta}{2}.
\end{equation}
For $N$ sensors, $U_\theta=e^{-i\theta\mathcal X/2}$ and the physical slope is
$G=i\langle[\mathcal X/2,\mathcal Y]\rangle=-\langle\mathcal Z\rangle$.
For $G\ne0$, error propagation gives
\begin{equation}
 (\Delta\theta)^2=\frac{(\Delta\mathcal Y)^2}{|\langle\mathcal Z\rangle|^2},
 \qquad
 \xi_R^2=\frac{(\Delta\theta)^2}{1/N}
 =\frac{N(\Delta\mathcal Y)^2}{|\langle\mathcal Z\rangle|^2},
\end{equation}
where the product reference has $(\Delta\theta)^2_{\rm SQL}=1/N$.
This is the conventional Wineland parameter with precisely the axes, response sign, and physical phase used in the main text.

\subsection{Two-component magnetic-field sensing}

The qutrit protocol in the main text estimates the two parameters encoded by
$u_{\boldsymbol\theta}=\exp[-i(\theta_xs_x+\theta_ys_y)]$.
For a pure reference, $f_{\alpha\beta}=4\Cov_\phi(s_\alpha,s_\beta)$, with $\alpha,\beta\in\{x,y\}$.
We minimize $C_q=\Tr(f^{-1})$ subject to full rank and weak commutativity, $\langle s_z\rangle_\phi=0$.

\textit{Qubit.}
Write $\ket\phi=\sin(\theta_0/2)e^{i\phi_0}\ket{-1/2}+\cos(\theta_0/2)\ket{1/2}$ and let $\boldsymbol n_\perp=\sin\theta_0(\cos\phi_0,\sin\phi_0)^T$.
The QFIM and its determinant are
\begin{equation}
 f=\mathbb I_2-\boldsymbol n_\perp\boldsymbol n_\perp^T,
 \qquad \det f=\cos^2\theta_0.
\end{equation}
For nonsingular $f$,
\begin{equation}
 C_q=\frac{2-\sin^2\theta_0}{1-\sin^2\theta_0}.
\end{equation}
Weak commutativity requires $\langle s_z\rangle_\phi=\frac12\cos\theta_0=0$, which makes $f^{-1}$ singular.
Thus a pure qubit reference cannot simultaneously provide two independent and weakly compatible channels.

\textit{Qutrit.}
A short analytic bound determines the optimum without a state parametrization.
Using $s_x^2+s_y^2+s_z^2=2\mathbb I_3$, any pure qutrit reference satisfies
\begin{equation}
 \Tr f=4\left(2-\langle s_z^2\rangle_\phi
 -\langle s_x\rangle_\phi^2-\langle s_y\rangle_\phi^2\right)\le8.
\end{equation}
For positive eigenvalues $\kappa_1,\kappa_2$ of $f$,
\begin{equation}
 C_q=\frac{1}{\kappa_1}+\frac{1}{\kappa_2}
 \ge\frac{4}{\kappa_1+\kappa_2}\ge\frac12.
 \label{eq-si-qutrit-cost-bound}
\end{equation}
The state $\ket0$ saturates both inequalities and obeys weak commutativity.
Conversely, saturation requires $\langle s_z^2\rangle=0$, which selects $\ket0$ up to a global phase.
Hence
\begin{equation}
 \ket\varphi=\ket0,\qquad f_{\rm op}=4\mathbb I_2,
 \qquad \Gamma_{\rm SQL}=\frac{\mathbb I_2}{4N},\qquad C_{\rm SQL}=\frac{1}{2N}.
\end{equation}
The corresponding tangent vectors are
\begin{equation}
 \ket{\nu_x}=\frac{\ket++\ket-}{\sqrt2},\qquad
 \ket{\nu_y}=\frac{i(\ket--\ket+)}{\sqrt2},\qquad
 \Omega_{\alpha\beta}=\delta_{\alpha\beta},
\end{equation}
where $\ket\pm=\ket{m_z=\pm1}$.
Thus $\Delta_x=\Delta_y=1$ and the physical signals equal the normalized channel quadratures, $X_x=s_x$ and $X_y=s_y$.
Their conjugate responses are
\begin{equation}
 Y_\alpha=-i\ket0\bra{\nu_\alpha}+i\ket{\nu_\alpha}\bra0,
 \qquad \alpha\in\{x,y\}.
 \label{eq-si-qutrit-response-quadratures}
\end{equation}
In spin notation these are $Y_x=-\{s_y,s_z\}$ and $Y_y=\{s_x,s_z\}$, with $Z_x=s_y^2-s_z^2$ and $Z_y=s_x^2-s_z^2$, as in the main text.
For the collective operators $\mathcal X_\alpha=\sum_jX_\alpha^{(j)}$ and $\mathcal Y_\alpha=\sum_jY_\alpha^{(j)}$, define
\begin{equation}
 \mathcal C_{\alpha\beta}=\Cov_\Psi(\mathcal Y_\alpha,\mathcal Y_\beta),\qquad
 G_{\beta\alpha}=\left.\frac{\partial\langle\mathcal Y_\beta\rangle}{\partial\theta_\alpha}\right|_0
 =i\langle[\mathcal X_\alpha,\mathcal Y_\beta]\rangle_\Psi.
 \label{eq-si-tranduction-2d}
\end{equation}
The physical phases already coincide with the normalized channel coordinates, so
\begin{equation}
 \Gamma_{\rm GW}=G^{-1}\mathcal C(G^T)^{-1}
 =(G^T\mathcal C^{-1}G)^{-1},\qquad
 \xi_{\rm GW}^2=2N\Tr\Gamma_{\rm GW}.
\end{equation}
For $\ket{\Psi_{\rm p}}=\ket0^{\otimes N}$, $\mathcal C=N\mathbb I_2$ and $G=-2N\mathbb I_2$; hence $\Gamma_{\rm GW}=\mathbb I_2/(4N)$ and $\xi_{\rm GW}^2=1$.

\subsection{Three-component magnetic-field sensing with a four-level qudit}

\begin{figure}[th]
    \centering
    \includegraphics[width=0.75\linewidth]{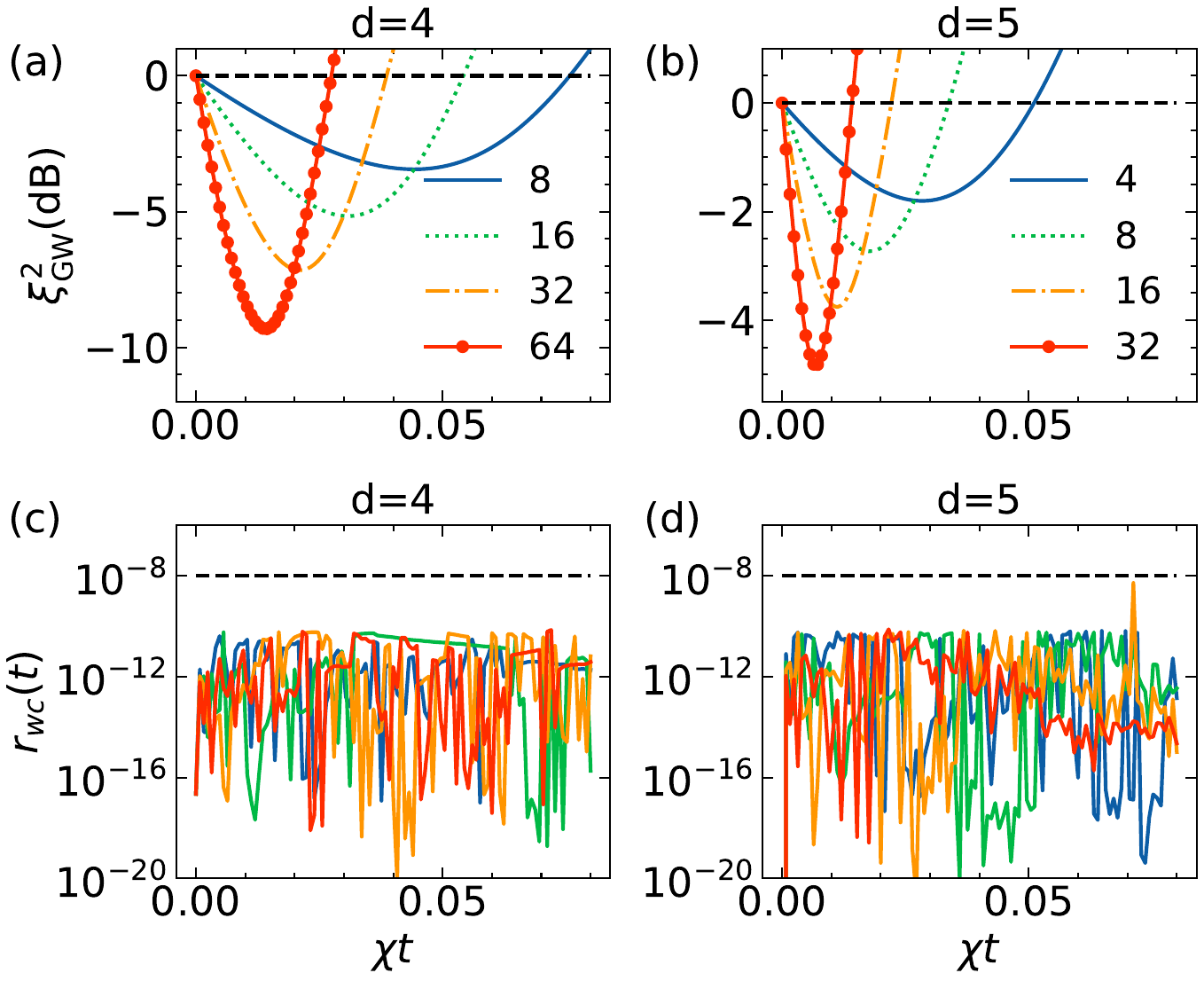}
    \caption{(a)-(b) Optimized squeezing parameters versus evolution time $\chi t$ for different system sizes $N$, with (a) $d=4$ and (b) $d=5$. The black dashed line marks the product-state SQL. (c)-(d) Weak-commutativity residual $r_{\rm wc}(t)=\max_{\alpha=x,y,z}|\langle L_\alpha\rangle_t|/N$, where $L_\alpha=\Lambda_\alpha$ for $d=4$ and $L_\alpha=S_\alpha$ for $d=5$. The black dashed line at $10^{-8}$ marks the numerical tolerance.}
    \label{fig-three-component-OAT}
\end{figure}

We next consider three-component sensing with a composite $d=4$ sensor made of two qubits. The single-site sensing geometry was discussed in Ref.~\cite{Isogawa2026Entanglement-Assisted}. The local Hilbert space is spanned by
\begin{equation}
    \ket{0}_a \ket{0}_b,\quad \ket{0}_a \ket{1}_b,\quad  \ket{1}_a \ket{0}_b,\quad \ket{1}_a \ket{1}_b.
\end{equation}
Qubit $a$ couples to the field through $H=\vec B\cdot\vec\sigma_a$, while qubit $b$ is an ancilla with negligible field coupling. Absorbing the coupling strength and interrogation time into the phases gives
\begin{equation}
    u_{\boldsymbol{\theta}} = \exp(-i \theta_x \sigma_a^x \otimes \mathbb{I}_b-i \theta_y \sigma_a^y \otimes \mathbb{I}_b-i \theta_z \sigma_a^z \otimes \mathbb{I}_b  )  = \exp(-i \theta_x \lambda_x - i\theta_y \lambda_y -i \theta_z \lambda_z ).
\end{equation}
For this sensing task and probe state $\ket{\phi} \in \mathcal{H}_4$, the weak commutativity condition is 
\begin{equation}
 \langle \sigma_a^x \otimes \mathbb{I}_b\rangle_\phi=\langle \sigma_a^y \otimes \mathbb{I}_b\rangle_\phi=\langle \sigma_a^z \otimes \mathbb{I}_b\rangle_\phi=0.
\end{equation}
We expand the qudit probe state as
\begin{equation}
    \ketp{\phi}  =
    z_{00}\ketp{0_a 0_b} +z_{01}\ketp{0_a 1_b} + z_{10}\ketp{1_a 0_b} + z_{11}\ketp{1_a 1_b},
\end{equation}
and define 
\begin{equation}
    \mathbf u=(z_{00},z_{01}),\qquad
    \mathbf v=(z_{10},z_{11}),
\end{equation}
with $\mathbf u \mathbf u^{\dagger} + \mathbf v \mathbf v^{\dagger} = 1$.
The reduced density matrix for qubit a can be written as
\begin{equation}
    \rho_a = \Tr_b\proj{\phi} = \frac{1}{2} \left( \mathbb{I}_a +\mathbf r\cdot\boldsymbol{\sigma}_a \right) =     \begin{pmatrix}
    \mathbf u\mathbf u^\dagger & \mathbf u\mathbf v^\dagger\\
    \mathbf v\mathbf u^\dagger & \mathbf v\mathbf v^\dagger
    \end{pmatrix},
\end{equation}
with $\mathbf{r} = (r_x, r_y, r_z) = ( \mathbf u \mathbf v^{\dagger} + \mathbf v \mathbf u^{\dagger}, i \mathbf u \mathbf v^{\dagger} - i \mathbf v \mathbf u^{\dagger},\mathbf u \mathbf u^{\dagger} - \mathbf v \mathbf v^{\dagger})$.
The weak-commutativity condition demands
\begin{equation}
    r_x = 0, \quad r_y = 0, \quad r_z = 0.
    \label{eq-si-weak-commutative-three-component}
\end{equation}
Thus the reduced state of qubit $a$ must be
\begin{equation}
    \rho_a = \frac{1}{2}\mathbb{I}_2
\end{equation}
For a pure state of two qubits, it is equivalent to maximal two-qubit entanglement. 
The general optimal reference state can therefore be written as
\begin{equation}
    \ketp{\varphi} = \frac{1}{\sqrt2}\left(\ketp{0}_a\ketp{\chi_0}_b +\ketp{1}_a\ketp{\chi_1}_b \right), \quad\langle\chi_m |\chi_n\rangle=\delta_{mn}.
\end{equation}
The Bell states 
\begin{equation}
    \ket{\Phi^{\pm}} = \frac{\ket{00} \pm \ket{11}}{\sqrt2}, \quad     \ket{\Psi^{\pm}} = \frac{\ket{01} \pm \ket{10}}{\sqrt2}.
\end{equation}
are convenient members of the optimal reference family for compatible three-component sensing.
The single-site QFIM is 
\begin{align}
    f_{\alpha\beta} = 
    4 \Cov_{\phi}(\lambda_\alpha,\lambda_\beta) & =  2\Tr_a\left[\rho_a\{\sigma_\alpha,\sigma_\beta\}\right]-4\Tr_a(\rho_a\sigma_\alpha)\Tr_a(\rho_a\sigma_\beta) \nonumber \\
    & = 4(\delta_{\alpha\beta}  - r_\alpha r_\beta) \nonumber \\
    & = 4 (\mathbb{I}_3 - \mathbf{r} \mathbf{r}^{T})_{\alpha\beta}.
\end{align}
The inverse of the single-site QFIM is 
\begin{equation}
    f^{-1} = \frac{1}{4}(\mathbb{I}_3 + \frac{\mathbf r\mathbf r^T}{1-|\mathbf r|^2}),
\end{equation}
Imposing Eq.~\eqref{eq-si-weak-commutative-three-component} gives 
\begin{equation}
    C_q = \Tr(f^{-1}) = 3/4.
\end{equation}
We choose $\ket{\varphi} = \ket{\Phi^{+}}$. The tangent vectors are given as 
\begin{equation}
    |\nu_x\rangle = (\lambda_x -\langle \lambda_x\rangle_{\varphi}) \ket{\varphi} = \ket{\Psi^{+}}, \quad  |\nu_y\rangle = (\lambda_y -\langle \lambda_y\rangle_{\varphi}) \ket{\varphi} = -i\ket{\Psi^{-}},\quad    |\nu_z\rangle = (\lambda_z -\langle \lambda_z\rangle_{\varphi}) \ket{\varphi} = \ket{\Phi^{-}},
\end{equation}
with $\Delta_\alpha=1$ and $\ket{\tilde{\nu}_\alpha} = \ket{\nu_\alpha}$.
The channel quadratures are defined as
\begin{equation}
X_\alpha=|\Phi^{+}\rangle\langle\tilde\nu_\alpha|+|\tilde\nu_\alpha\rangle\langle\Phi^{+}|,\quad
 Y_\alpha=-i|\Phi^{+}\rangle\langle\tilde\nu_\alpha|+i|\tilde\nu_\alpha\rangle\langle\Phi^{+}|, \quad Z_\alpha = \ket{\Phi^{+}} \bra{\Phi^{+}} - \ket{\tilde{\nu}_\alpha} \bra{\tilde{\nu}_\alpha}
\end{equation}
The physical generators and channel quadratures have the same tangent
action,
\begin{equation}
\lambda_\alpha|\Phi^+\rangle = X_\alpha|\Phi^+\rangle = |\nu_\alpha\rangle,
\end{equation}
but they are not identical operators on $\mathcal H_4$. Instead,
\begin{equation}
\lambda_\alpha=X_\alpha+R_\alpha, \qquad R_\alpha=Q\lambda_\alpha Q\neq0, \qquad Q=\mathbb I-|\Phi^+\rangle\langle\Phi^+|.
\end{equation}
Explicitly,
\begin{align}
R_x &= -|\Phi^-\rangle\langle\Psi^-| -|\Psi^-\rangle\langle\Phi^-|, \nonumber\\
R_y &= i|\Psi^+\rangle\langle\Phi^-| -i|\Phi^-\rangle\langle\Psi^+|, \nonumber\\
R_z &= |\Psi^+\rangle\langle\Psi^-| +|\Psi^-\rangle\langle\Psi^+|.
\end{align}
Equivalently,
\begin{align}
R_x&=\frac12\left(\sigma_a^x\otimes\mathbb I_b -\mathbb I_a\otimes\sigma_b^x\right), \nonumber\\
R_y&=\frac12 \left(\sigma_a^y\otimes\mathbb I_b +\mathbb I_a\otimes\sigma_b^y\right),\nonumber\\
R_z&=\frac12\left(\sigma_a^z\otimes\mathbb I_b-\mathbb I_a\otimes\sigma_b^z\right).
\end{align}
These operators act entirely within the subspace orthogonal to
$|\Phi^+\rangle$ and satisfy
\begin{equation}
R_\alpha|\Phi^+\rangle=0, \qquad \lambda_\alpha=X_\alpha+R_\alpha.
\end{equation}
For an ensemble, we retain these task-selected responses and the full physical generators. 
The collective physical generators and response quadratures are $\Lambda_\alpha=\sum_j\lambda_\alpha^{(j)}$ and $\mathcal Y_\alpha=\sum_jY_\alpha^{(j)}$, respectively. Their covariance and transduction matrices are
\begin{equation}
 \mathcal C_{\alpha\beta}=\Cov(\mathcal Y_\alpha,\mathcal Y_\beta),\qquad
G_{\beta\alpha}=\left.\frac{\partial\langle\mathcal Y_\beta\rangle}{\partial\theta_\alpha}\right|_0 =i\langle[\Lambda_\alpha,\mathcal Y_\beta]\rangle.
 \label{eq-si-tranduction-3d}
\end{equation}
For nonsingular $G$, the ideal collective-response sensitivity is 
\begin{equation}
 \Gamma_{\rm GW}=(G^T\mathcal C^{-1}G)^{-1}
 =G^{-1}\mathcal C(G^T)^{-1}
\end{equation}
For the product reference state,
\begin{equation}
\mathcal C=N\mathbb I_3,\qquad G=-2N\mathbb I_3,\qquad\Gamma_{\rm GW}=\Gamma_{\rm SQL}=\frac{\mathbb I_3}{4N},
\end{equation}
and therefore
\begin{equation}
\xi_{\rm GW}^2 =\frac{4N}{3}\Tr\Gamma_{\rm GW}
\end{equation}
Consider the one-axis-twisting-like interaction
\begin{equation}
    H_{\rm OAT} = \chi (\Lambda_x^2 + \Lambda_y^2 + \Lambda_z^2),
\end{equation}
with $\Lambda_{\alpha} = \sum_j (\sigma_a^\alpha \otimes \mathbb{I}_b)^{(j)}$.
Starting from the product state $\ket{\Phi^{+}}^{\otimes N}$, the dynamics at different system sizes are presented in Fig.~\ref{fig-three-component-OAT}(a) and (c).
Before evaluating the squeezing parameter, we optimize a collective $\mathrm{SU}(4)$ rotation, applied identically to every qudit and constrained to preserve weak commutativity, to align the generated state with the measured responses.
The optimized curves show beyond-SQL gain over the simulated sizes. 

\subsection{Three-component magnetic-field sensing in a five-level qudit}

We next consider three-component magnetic-field sensing with a spin-$2$ system, corresponding to a five-level qudit. The sensing unitary is
\begin{equation}
u_{\boldsymbol{\theta}} =
\exp\left[ -i\left( \theta_x s_x+\theta_y s_y+\theta_z s_z\right)\right],
\end{equation}
where $s_\alpha$ are the spin-$2$ generators. For a general single-site probe
\begin{equation}
|\phi\rangle=\sum_{m=-2}^{2}z_m|m\rangle,\qquad \sum_{m=-2}^{2}|z_m|^2=1,
\end{equation}
the single-site QFIM is
\begin{equation}
f_{\alpha\beta}(\phi)=4\Cov_{\phi}(s_\alpha,s_\beta),\qquad\alpha,\beta\in\{x,y,z\}.
\end{equation}
Because the spin generators satisfy
\begin{equation}
[s_\alpha,s_\beta] =i\sum_{\gamma}\epsilon_{\alpha\beta\gamma}s_\gamma,
\end{equation}
the weak-commutativity conditions reduce to
\begin{equation}
\langle s_x\rangle_\phi=\langle s_y\rangle_\phi =\langle s_z\rangle_\phi=0.
\label{eq-si-weak-commutative-d5}
\end{equation}
Under these constraints, the QFIM becomes
\begin{equation}
f_{\alpha\beta}=2\langle\{s_\alpha,s_\beta\}\rangle_\phi .
\end{equation}
The optimization can be performed analytically. For local spin $s=2$,
\begin{equation}
s_x^2+s_y^2+s_z^2 = s(s+1)\mathbb I = 6\mathbb I.
\end{equation}
Therefore, for any state satisfying Eq.~\eqref{eq-si-weak-commutative-d5},
\begin{equation}
\Tr f = 4\sum_{\alpha=x,y,z}\langle s_\alpha^2\rangle_\phi =4s(s+1) = 24.
\end{equation}
Let $\kappa_1,\kappa_2,\kappa_3>0$ denote the eigenvalues of $f$.
The equal-weight cost obeys
\begin{equation}
C_q = \Tr(f^{-1})=\sum_{\mu=1}^{3}\frac{1}{\kappa_\mu}\geq\frac{9}{\sum_{\mu=1}^{3}\kappa_\mu}=\frac{3}{8},
\label{eq-si-d5-cost-bound}
\end{equation}
where equality holds if and only if
\begin{equation}
f_{\rm op}=8\mathbb I_3.
\end{equation}
Thus, the optimum corresponds to an isotropic spin covariance matrix.
One state that saturates this bound is
\begin{equation}
|\varphi\rangle = \frac{1}{\sqrt{2}}|0\rangle + \frac{i}{2}\left(|-2\rangle+|+2\rangle\right).
\label{eq-si-wf-d5-single-site}
\end{equation}
For this state,
\begin{equation}
\langle s_\alpha\rangle_\varphi=0,\qquad\frac{1}{2} \langle\{s_\alpha,s_\beta\}\rangle_\varphi=2\delta_{\alpha\beta},
\end{equation}
and hence
\begin{equation}
f_{\rm op}=8\mathbb I_3,
\qquad
C_q=\frac{3}{8},
\qquad
\Gamma_{\rm SQL}=\frac{\mathbb I_3}{8N},
\qquad
C_{\rm SQL}=\Tr\Gamma_{\rm SQL}=\frac{3}{8N}.
\label{eq-si-d5-sql}
\end{equation}

We next construct the tangent-space quadratures associated with this optimal reference state.
Since $\langle s_\alpha\rangle_\varphi=0$, the tangent vectors are simply
\begin{equation}
|\nu_\alpha\rangle = s_\alpha|\varphi\rangle.
\end{equation}
Explicitly,
\begin{align}
|\nu_x\rangle =e^{i\pi/6}\left(|+1\rangle+|-1\rangle\right),
\qquad
|\nu_y\rangle =e^{i\pi/3}\left(|-1\rangle-|+1\rangle\right),
\qquad
|\nu_z\rangle =i\left(|+2\rangle-|-2\rangle\right).
\end{align}
They obey
\begin{equation}
\Omega_{\alpha\beta}
=
\langle\nu_\alpha|\nu_\beta\rangle
=
2\delta_{\alpha\beta}.
\end{equation}
Therefore,
\begin{equation}
\Delta_\alpha=(\Delta s_\alpha)_\varphi=\sqrt{2}, \qquad|\tilde{\nu}_\alpha\rangle =\frac{|\nu_\alpha\rangle}{\sqrt{2}},
\end{equation}
and the three normalized tangents form an orthonormal basis of the sensing subspace.

As in the preceding examples, we define the local channel quadratures
\begin{align}
X_\alpha = |\varphi\rangle\langle\tilde{\nu}_\alpha| + |\tilde{\nu}_\alpha\rangle\langle\varphi|,\qquad Y_\alpha =-i|\varphi\rangle\langle\tilde{\nu}_\alpha| + i|\tilde{\nu}_\alpha\rangle\langle\varphi|, \qquad Z_\alpha = |\varphi\rangle\langle\varphi| - |\tilde{\nu}_\alpha\rangle\langle\tilde{\nu}_\alpha|.
\end{align}
The physical spin generators and normalized signal quadratures have the same tangent action up to the tangent norm,
\begin{equation}
s_\alpha|\varphi\rangle =\sqrt{2} X_\alpha|\varphi\rangle =|\nu_\alpha\rangle.
\end{equation}
Equivalently,
\begin{equation}
s_\alpha =\sqrt{2}X_\alpha+R_\alpha,\qquad R_\alpha = Q s_\alpha Q, \qquad Q=\mathbb I-|\varphi\rangle\langle\varphi|,
\end{equation}
where $R_\alpha|\varphi\rangle=0$.
To express these residual operators explicitly, we introduce the fourth state in the orthogonal subspace,
\begin{equation}
|\chi\rangle = \frac{1}{\sqrt{2}}|0\rangle-\frac{i}{2}\left(|2\rangle+|-2\rangle\right),
\end{equation}
such that
$\{|\tilde\nu_x\rangle,|\tilde\nu_y\rangle,
|\tilde\nu_z\rangle,|\chi\rangle\}$ forms an orthonormal basis of the $Q$ subspace. 
In this basis, the residual generators take the compact form
\begin{align}
R_x &= \sqrt{2}e^{-i\pi/3}|\tilde\nu_x\rangle\langle\chi|+e^{-i5\pi/6}|\tilde\nu_y\rangle\langle\tilde\nu_z|+\mathrm{H.c.}, \nonumber\\
R_y &= e^{i5\pi/6} |\tilde\nu_x\rangle\langle\tilde\nu_z| + \sqrt{2}e^{i\pi/3} |\tilde\nu_y\rangle\langle\chi| +\mathrm{H.c.},\nonumber\\
R_z &= e^{-i5\pi/6}|\tilde\nu_x\rangle\langle\tilde\nu_y|-\sqrt{2}|\tilde\nu_z\rangle\langle\chi|+\mathrm{H.c.}.
\label{eq-si-d5-residual-generators}
\end{align}
For an ensemble of $N$ sensors, we introduce the collective physical generators and response quadratures
\begin{equation}
S_\alpha = \sum_{j=1}^{N}s_\alpha^{(j)},\qquad \mathcal Y_\alpha =\sum_{j=1}^{N}Y_\alpha^{(j)}.
\end{equation}
Their covariance and transduction matrices are
\begin{equation}
\mathcal C_{\alpha\beta} = \Cov(\mathcal Y_\alpha,\mathcal Y_\beta),\qquad G_{\beta\alpha}=\left.\frac{\partial\langle \mathcal Y_\beta\rangle}{\partial\theta_\alpha}\right|_{\boldsymbol{\theta}=0}=i\langle[ S_\alpha,\mathcal Y_\beta]\rangle.
\label{eq-si-transduction-d5}
\end{equation}
The generalized-Wineland sensitivity matrix for the physical field parameters is
\begin{equation}
\Gamma_{\rm GW}=(G^T\mathcal C^{-1}G)^{-1}= G^{-1}\mathcal C(G^T)^{-1}.
\end{equation}
Using the SQL in Eq.~\eqref{eq-si-d5-sql}, we define
\begin{equation}
\xi_{\rm GW}^2=\frac{\Tr\Gamma_{\rm GW}}{C_{\rm SQL}}=\frac{8N}{3} \Tr\Gamma_{\rm GW}.
\label{eq-si-d5-gw}
\end{equation}
Consider the one-axis-twisting-like interaction
\begin{equation}
    H_{\rm OAT} = \chi (S_x^2 + S_y^2 + S_z^2),
\end{equation}
Starting from the product state $\ket{\varphi}^{\otimes N}$, the dynamics at different system sizes are presented in Fig.~\ref{fig-three-component-OAT}(b) and (d).
Before evaluating the squeezing parameter, we optimize a collective $\mathrm{SU}(5)$ rotation, applied identically to every qudit and constrained to preserve weak commutativity, to align the generated state with the measured responses.
The optimized curves show beyond-SQL gain over the simulated sizes. 

\section{Quadrature-algebra and truncated Wigner analysis of twisting-induced multiparameter squeezing}

We use the channel-quadrature algebra to show how the TAT-like Hamiltonian in the main text generates multiparameter squeezing.
The same leading-order construction is the quadrature form of the truncated Wigner approximation (TWA) \cite{Polkovnikov2010Phase, Hosseinabadi2025User-Friendly}.

We consider two compatible sensing channels around the optimal product state $|\Psi_{\rm p}\rangle=|0\rangle^{\otimes N}$.
Weak commutativity makes the tangent-vector Gram matrix real; for the optimal qutrit reference used here, the normalized tangents are additionally orthonormal, $\langle\tilde\nu_\alpha|\tilde\nu_\beta\rangle=\delta_{\alpha\beta}$.
The local quadratures $X_\alpha$, $Y_\alpha$, and $Z_\alpha$ follow Eq.~\eqref{eq-si-channel-quadratures} with $|\varphi\rangle=|0\rangle$.
To also describe commutators between different channels, define
\begin{align}
 [X_\alpha,Y_\beta]&=2iZ_{\alpha\beta},\quad 
 [X_\alpha,X_\beta]=[Y_\alpha,Y_\beta] =|\tilde\nu_\alpha\rangle\langle\tilde\nu_\beta|-|\tilde\nu_\beta\rangle\langle\tilde\nu_\alpha|.
 \label{eq-si-quadrature-algebra}
\end{align}
Here
\begin{equation}
 Z_{\alpha\beta}=\delta_{\alpha\beta}|\varphi\rangle\langle\varphi|
 -\frac{|\tilde\nu_\alpha\rangle\langle\tilde\nu_\beta|
       +|\tilde\nu_\beta\rangle\langle\tilde\nu_\alpha|}{2},
 \qquad
 \mathcal Z_{\alpha\beta}=\sum_j Z_{\alpha\beta}^{(j)},
\end{equation}
so $Z_{\alpha\alpha}=Z_\alpha$ and $\mathcal Z_{\alpha\alpha}=\mathcal Z_\alpha$.
On the initial state $\ket{\Psi_{\rm p}}$, the tangent-space coherences vanish and
\begin{align}
 \langle\mathcal Z_{\alpha\beta}\rangle_{\Psi_{\rm p}}&=N\delta_{\alpha\beta},\nonumber\\
 \Cov_{\Psi_{\rm p}}(\mathcal X_\alpha,\mathcal X_\beta)&=\Cov_{\Psi_{\rm p}}(\mathcal Y_\alpha,\mathcal Y_\beta) =N\delta_{\alpha\beta},\nonumber\\
 \Cov_{\Psi_{\rm p}}(\mathcal X_\alpha,\mathcal Y_\beta)&=0.
 \label{eq-si-initial-quadrature-moments}
\end{align}
We introduce the fluctuation quadratures
\begin{equation}
 q_\alpha=\frac{\mathcal X_\alpha}{\sqrt{2N}},\qquad
 p_\alpha=\frac{\mathcal Y_\alpha}{\sqrt{2N}}.
 \label{eq-si-rescaled-quadratures}
\end{equation}
The TAT-like Hamiltonian becomes
\begin{equation}
H_{\rm TAT} =\chi\sum_{\alpha=x,y} \{\mathcal X_\alpha,\mathcal Y_\alpha \} =\tilde{\chi}\sum_{\alpha=x,y}
\{q_\alpha,p_\alpha\},
\qquad
\tilde{\chi}=2\chi N.
\label{eq-si-general-tat-hamiltonian}
\end{equation}
The exact Heisenberg equations are
\begin{align}
\dot q_\alpha &=-i\tilde{\chi}\sum_{\beta=x,y}\left( \{[q_\alpha,q_\beta],p_\beta\} +\{[q_\alpha,p_\beta],q_\beta\} \right), \nonumber\\
\dot p_\alpha &=-i\tilde{\chi}\sum_{\beta=x,y}
\left( \{[p_\alpha,q_\beta],p_\beta\} +\{[p_\alpha,p_\beta],q_\beta\} \right).
\label{eq-si-exact-quadrature-equations}
\end{align}
In the Gaussian large-$N$ limit underlying the TWA, the initial collective
fluctuations are represented by
\begin{equation}
W_{\rm p}(\boldsymbol q,\boldsymbol p) =\frac{1}{\pi^2}\exp\left[ -\sum_{\alpha=x,y}(q_\alpha^2+p_\alpha^2) \right],
\label{eq-si-quadrature-wigner-function}
\end{equation}
so the initial variables satisfy
\begin{equation}
\overline{q_\alpha q_\beta} =\overline{p_\alpha p_\beta} =\frac{\delta_{\alpha\beta}}{2}, \quad\overline{q_\alpha p_\beta}=0.
\end{equation}
Near the product state,
$[q_\alpha,p_\beta]\simeq i\delta_{\alpha\beta}$, while the cross-channel commutators are subleading in $1/\sqrt N$.
For $|\tilde{\chi}t|\ll1$, the linearized equations therefore reduce to
\begin{equation}
\dot q_\alpha\simeq2\tilde{\chi}q_\alpha,
\qquad
\dot p_\alpha\simeq-2\tilde{\chi}p_\alpha.
\label{eq-si-quadrature-heisenberg-equations}
\end{equation}
Thus the signal fluctuations are amplified while the conjugate response
fluctuations are squeezed. To leading order,
\begin{align}
\Cov(\mathcal X_\alpha,\mathcal X_\beta) &\simeq N(1+8\chi Nt)\delta_{\alpha\beta},
\nonumber\\
\mathcal C_{\alpha\beta} =\Cov(\mathcal Y_\alpha,\mathcal Y_\beta) &\simeq N(1-8\chi Nt)\delta_{\alpha\beta},
\nonumber\\
G_{\beta\alpha} =i\langle[\mathcal X_\alpha,\mathcal Y_\beta]\rangle &\simeq-2N\delta_{\alpha\beta}.
\label{eq-si-quadrature-leading-covariances}
\end{align}
For the two qutrit channels ($k=2$), the generalized Wineland parameter is therefore
\begin{align}
\xi_{\rm GW}^2 &=
2N\Tr\left[ G^{-1}\mathcal C (G^T)^{-1} \right] \nonumber \\
&\simeq1-8\chi Nt<1.
\end{align}
Hence the qutrit TAT interaction squeezes both response quadratures at short times in this large-$N$ approximation. The same linearization method can be applied to one-axis-twisting candidates, but the resulting squeezing directions and gain must be determined for each Hamiltonian.

\section{Conservation of weak commutativity under twisting dynamics}

For the qutrit magnetometry example, weak commutativity requires $\langle S_z\rangle=0$. The channel quadratures satisfy
\begin{equation}
 [S_z,\mathcal X_x]=i\mathcal X_y,\quad [S_z,\mathcal X_y]=-i\mathcal X_x,\quad
 [S_z,\mathcal Y_x]=i\mathcal Y_y,\quad [S_z,\mathcal Y_y]=-i\mathcal Y_x.
\end{equation}
Thus both pairs transform as vectors under rotations generated by $S_z$. For the TAT-like Hamiltonian defined in the main text,
\begin{equation}
 H_{\rm TAT}=\chi\sum_{\alpha=x,y}\{\mathcal X_\alpha,\mathcal Y_\alpha\},
\end{equation}
one finds
\begin{equation}
 [S_z,H_{\rm TAT}]=i\chi\bigl(\{\mathcal X_y,\mathcal Y_x\}+\{\mathcal X_x,\mathcal Y_y\}-\{\mathcal X_x,\mathcal Y_y\}-\{\mathcal X_y,\mathcal Y_x\}\bigr)=0.
\end{equation}
The OAT-like Hamiltonian obeys the same symmetry,
\begin{equation}
 H_{\rm OAT}=\chi\sum_{\alpha=x,y}\mathcal X_\alpha^2,
 \qquad [S_z,H_{\rm OAT}]=0.
\end{equation}
Because the optimal product state $|\Psi_{\rm p}\rangle=|0\rangle^{\otimes N}$ has $\langle S_z\rangle=0$, either evolution preserves $\langle S_z\rangle=0$ and hence the weak-commutativity condition.

\section{Naimark implementation of the fixed qutrit POVM}
\label{sec-si-naimark-readout}

The main text defines the fixed six-outcome POVM $\{\Pi_{x,+},\Pi_{x,-},\Pi_{y,+},\Pi_{y,-},\Pi_+,\Pi_-\}$, with
\begin{equation}
 \ket{\varsigma_{\alpha,\pm}}=\frac{\ket0\pm i\ket{\nu_\alpha}}{\sqrt2},
 \qquad
 \Pi_{\alpha,\pm}=\frac12\ket{\varsigma_{\alpha,\pm}}\bra{\varsigma_{\alpha,\pm}},
 \qquad
 \Pi_\pm=\frac12\ket\pm\bra\pm.
\end{equation}
The single-site score $\tilde y_\alpha(m)$ is $\pm2$ for outcome $m=(\alpha,\pm)$ and zero for the other four outcomes.
Here we give an explicit Naimark dilation of that POVM; the statistical properties of its direct-product extension are derived in the next section.
As an illustrative atomic realization, we identify the qutrit basis with an $F=1$ manifold as $|+\rangle\equiv|1,+1\rangle$, $|0\rangle\equiv|1,0\rangle$, and $|-\rangle\equiv|1,-1\rangle$, and use an $F=2$ manifold as the ancilla. These hyperfine labels apply only to this dilation example. The trapped-ion encoding in the last End Matter instead uses $|0\rangle=|F=0,m_F=0\rangle$ and $|\pm\rangle=|F=1,m_F=\pm1\rangle$ \cite{Senko2015Realization}. The abstract qutrit operators and POVM remain the same, but implementing the six-outcome dilation in that ion requires a suitable enlarged detection space or ancilla, rather than the $F=1\oplus F=2$ level assignment used below.
Assign the four response outcomes to
\begin{equation}
|a_{x,+}\rangle\equiv |2,-2\rangle,\quad
|a_{x,-}\rangle\equiv |2,-1\rangle,\quad
|a_{y,-}\rangle\equiv |2,0\rangle,\quad
|a_{y,+}\rangle\equiv |2,+1\rangle,
\end{equation}
and the remaining two outcomes to
\begin{equation}
|b_+\rangle\equiv |1,+1\rangle,\qquad
|b_-\rangle\equiv |1,-1\rangle.
\end{equation}
The corresponding isometry is
\begin{equation}
V= \frac{1}{\sqrt{2}}\left(\sum_{\alpha=x,y}\sum_{r=\pm}|a_{\alpha,r}\rangle\langle\varsigma_{\alpha,r}|+|b_+\rangle\langle+|+|b_-\rangle\langle-|\right),
\label{eq-si-naimark-isometry}
\end{equation}
which satisfies
\begin{equation}
V^\dagger V=\mathbb{I}_{F=1}.
\end{equation}
Any unitary $U_{\rm N}$ on $F=1\oplus F=2$ whose restriction to the $F=1$ subspace equals $V$ implements the desired POVM. Indeed,
\begin{equation}
p(a_{\alpha,r})=\frac12\langle\varsigma_{\alpha,r}|\rho|\varsigma_{\alpha,r}\rangle,\qquad p(b_r)=\frac12\langle r|\rho|r\rangle,
\end{equation}
so the Zeeman-resolved histogram gives $\langle Y_\alpha\rangle=2[p(a_{\alpha,+})-p(a_{\alpha,-})]$ for both $\alpha\in\{x,y\}$.

We denote the local dilation unitary by $U_{\rm N}$ to distinguish it from the many-body interaction-based readout $U_{\rm rd}$ applied before the POVM. One implementation of the dilation is obtained by decomposing
\begin{equation}
U_{\rm N}=W U_A.
\end{equation}
The first step performs partial transfer between the two hyperfine manifolds through the three $\Delta m_F=0$ couplings
\begin{equation}
|1,+1\rangle \leftrightarrow |2,+1\rangle,\qquad|1,0\rangle \leftrightarrow |2,0\rangle,\qquad|1,-1\rangle \leftrightarrow |2,-1\rangle,
\end{equation}
which can be implemented either by direct microwave hyperfine control or by an off-resonant Raman process. The effective Hamiltonian, with $\hbar$ restored, is
\begin{equation}
H_A= \frac{\hbar}{2} \sum_{m=-1}^{+1} \Omega_m e^{i\varphi_m}|2,m\rangle\langle 1,m| +\mathrm{h.c.}
\end{equation}
Choosing pulse areas
\begin{equation}
\Omega_{+1}t=\Omega_{-1}t=\frac{\pi}{2},\qquad \Omega_0 t=\pi,
\end{equation}
and phases $\varphi_m=0$, one obtains
\begin{align}
|1,+1\rangle &\xrightarrow{U_A}\frac{1}{\sqrt{2}}|1,+1\rangle-\frac{i}{\sqrt{2}}|2,+1\rangle,\\|1,0\rangle &\xrightarrow{U_A} -i|2,0\rangle,\\|1,-1\rangle &\xrightarrow{U_A}\frac{1}{\sqrt{2}}|1,-1\rangle-\frac{i}{\sqrt{2}}|2,-1\rangle.
\end{align}
The second step is a fixed unitary acting only within the $F=2$ manifold. Define
\begin{align}
|\chi_+\rangle &=
\frac{1}{2}\Bigl(|2,-2\rangle+|2,-1\rangle-|2,0\rangle+|2,+1\rangle\Bigr),\\
|\chi_0\rangle &=
\frac{1}{2}\Bigl(i|2,-2\rangle-i|2,-1\rangle+|2,0\rangle+|2,+1\rangle\Bigr),\\
|\chi_-\rangle &=
\frac{1}{2}\Bigl(|2,-2\rangle+|2,-1\rangle+|2,0\rangle-|2,+1\rangle\Bigr),
\end{align}
and
\begin{equation}
|\chi_d\rangle=
\frac{1}{2}\Bigl(-i|2,-2\rangle+i|2,-1\rangle+|2,0\rangle+|2,+1\rangle\Bigr).
\end{equation}
Then one may choose $W$ such that
\begin{equation}
W|2,+1\rangle=i|\chi_+\rangle,\qquad
W|2,0\rangle=i|\chi_0\rangle,\qquad
W|2,-1\rangle=i|\chi_-\rangle,
\end{equation}
with, for example,
\begin{equation}
W|2,-2\rangle=|\chi_d\rangle,\qquad
W|2,+2\rangle=|2,+2\rangle.
\end{equation}
For the explicit decomposition above, the restriction of $U_{\rm N}=WU_A$ to the input qutrit subspace is
\begin{align}
V_{\rm impl} =\frac{1}{\sqrt2}\bigl(
&i|a_{x,+}\rangle\langle\varsigma_{x,+}| -i|a_{x,-}\rangle\langle\varsigma_{x,-}|
\nonumber\\
&+|a_{y,-}\rangle\langle\varsigma_{y,-}|+|a_{y,+}\rangle\langle\varsigma_{y,+}|+|b_+\rangle\langle+|+|b_-\rangle\langle-|\bigr).
\end{align}
This differs from the target isometry $V$ only by outcome-dependent phases on the orthogonal readout states.
Such phases leave the induced POVM effects unchanged, so $U_{\rm N}$ implements the six-outcome POVM exactly.

Experimentally, $W$ can be synthesized by RF or microwave control within the $F=2$ manifold,
\begin{equation}
H_B(t)= \frac{\hbar}{2} \sum_{m=-2}^{+1}
\Omega^{(B)}_{m,m+1}(t)e^{i\varphi_m^{(B)}}
|2,m+1\rangle\langle 2,m| +\mathrm{h.c.},
\end{equation}
or, equivalently, by Raman couplings if an all-optical implementation is preferred. Since $W$ is fixed, it can be calibrated once and then reused for all measurements.

After $U_A$ and $W$, a single Zeeman-resolved projective measurement therefore
returns both response estimates. The unused channels $|1,0\rangle$ and
$|2,+2\rangle$ are ideally dark and provide direct diagnostics of leakage and
calibration errors.

\section{Added noise of the direct-product POVM}
\label{sec-si-product-povm-noise}

The main text states the added-noise penalty for the fixed local readout.
Here we supply its many-body second-moment derivation. For the single-site scores $\tilde y_\alpha(m)$ defined there,
\begin{equation}
 \sum_m \tilde y_\alpha(m)\Pi_m=Y_\alpha, \qquad \alpha\in\{x,y\}.
 \label{eq-first-moment-identity}
\end{equation}
On $N$ sensors, let
\begin{equation}
 \mathcal E_{\boldsymbol m}=\Pi_{m_1}^{(1)}\otimes\cdots\otimes\Pi_{m_N}^{(N)},\qquad\mathcal P_{\boldsymbol m}=\Tr(\rho_\Psi\mathcal E_{\boldsymbol m}),
 \label{eq-product-povm}
\end{equation}
where $\boldsymbol m=(m_1,\ldots,m_N)$, and define the estimator
\begin{equation}
 \widehat{\mathcal Y}_\alpha(\boldsymbol m) =\sum_{j=1}^{N}\tilde y_\alpha(m_j).
\end{equation}
Equation~\eqref{eq-first-moment-identity} gives $\mathbb E(\widehat{\mathcal Y}_\alpha) =\langle\mathcal Y_\alpha\rangle$, including for correlated probes, so the POVM record reproduces the main-text transduction matrix $G_{\beta\alpha} =\left.\partial_{\theta_\alpha}\langle\mathcal Y_\beta\rangle\right|_{\boldsymbol\theta=0}$.
To expose the additional noise, define
\begin{equation}
 M_{\alpha\beta} =\sum_m\tilde y_\alpha(m)\tilde y_\beta(m)\Pi_m.
\end{equation}
The classical second moment separates into intersite and onsite terms,
\begin{equation}
 \mathbb E \left(\widehat{\mathcal Y}_\alpha \widehat{\mathcal Y}_\beta\right)=\sum_{j\ne k}\langle Y_\alpha^{(j)}Y_\beta^{(k)}\rangle_\Psi +\sum_j\langle M_{\alpha\beta}^{(j)}\rangle_\Psi.
 \label{eq-si-povm-second-moment}
\end{equation}
The intersite contribution is identical to that of the collective operators;
only the onsite term differs. Consequently,
\begin{align}
 \mathcal C^{\rm POVM}_{\alpha\beta}&\equiv\Cov\!\left( \widehat{\mathcal Y}_\alpha,\widehat{\mathcal Y}_\beta\right)=\mathcal C_{\alpha\beta}+\mathcal Q_{\alpha\beta},
 \quad 
 \mathcal Q_{\alpha\beta} =\sum_{j=1}^{N}\langle\mathcal D_{\alpha\beta}^{(j)}\rangle_\Psi,
 \qquad \mathcal D_{\alpha\beta} =M_{\alpha\beta}-\frac12\{Y_\alpha,Y_\beta\}.
 \label{eq-output-covariance}
\end{align}
This is the covariance decomposition quoted in the main text. We now explicitly derive the local moments and the state-independent matrix bound.

\textit{Step 1: Local score second moments.}
The qutrit states $\ket0,\ket{\nu_x},\ket{\nu_y}$ form an orthonormal basis.
Define $P_0=\ket0\bra0$ and $P_\alpha=\ket{\nu_\alpha}\bra{\nu_\alpha}$, so that
$P_0+P_x+P_y=\mathbb I_3$.
Using $Y_\alpha=-i\ket0\bra{\nu_\alpha}+i\ket{\nu_\alpha}\bra0$, direct multiplication gives
\begin{equation*}
 Y_\alpha Y_\beta
 =\delta_{\alpha\beta}P_0+\ket{\nu_\alpha}\bra{\nu_\beta},
 \qquad Y_\alpha^2=P_0+P_\alpha.
\end{equation*}
The two nonzero scores for channel $\alpha$ are $+2$ and $-2$, and the corresponding POVM effects satisfy
\begin{equation*}
 \Pi_{\alpha,+}+\Pi_{\alpha,-}=\frac12(P_0+P_\alpha),
 \qquad
 M_{\alpha\alpha}
 =4(\Pi_{\alpha,+}+\Pi_{\alpha,-})
 =2(P_0+P_\alpha)=2Y_\alpha^2.
\end{equation*}
For $\alpha\ne\beta$, the two scores are never simultaneously nonzero on the same outcome, so $M_{\alpha\beta}=0$.
Substituting these moments into the definition of $\mathcal D_{\alpha\beta}$ yields
\begin{equation}
 M_{\alpha\beta}=2\delta_{\alpha\beta}Y_\alpha^2,
 \qquad
 \mathcal D_{\alpha\beta} =2\delta_{\alpha\beta}Y_\alpha^2-\frac12\{Y_\alpha,Y_\beta\}.
 \label{eq-si-qutrit-povm-onsite-noise}
\end{equation}

\textit{Step 2: Quadratic form of the local added noise.}
The quadrature products above give
\begin{equation*}
 \mathcal D_{xx}=P_0+P_x,\qquad
 \mathcal D_{yy}=P_0+P_y,\qquad
 \mathcal D_{xy}=\mathcal D_{yx}
 =-\frac12\left(\ket{\nu_x}\bra{\nu_y}+\ket{\nu_y}\bra{\nu_x}\right).
\end{equation*}
For any real $\boldsymbol v=(v_x,v_y)^T$, let
$\mathcal D(\boldsymbol v)=\sum_{\alpha,\beta}v_\alpha\mathcal D_{\alpha\beta}v_\beta$.
Combining the two off-diagonal terms, we obtain
\begin{equation*}
 \mathcal D(\boldsymbol v)
 =(v_x^2+v_y^2)P_0+v_x^2P_x+v_y^2P_y
 -v_xv_y\left(\ket{\nu_x}\bra{\nu_y}+\ket{\nu_y}\bra{\nu_x}\right).
\end{equation*}
The last three terms form a single outer product. Thus
\begin{equation}
 \sum_{\alpha,\beta}v_\alpha\mathcal D_{\alpha\beta}v_\beta
 =|\boldsymbol v|^2\ket0\bra0+\ket{w_{\boldsymbol v}}\bra{w_{\boldsymbol v}},
 \qquad
 \ket{w_{\boldsymbol v}}=v_x\ket{\nu_x}-v_y\ket{\nu_y}.
 \label{eq-si-qutrit-noise-quadratic-form}
\end{equation}

\textit{Step 3: Local operator bound.}
Orthogonality of the qutrit basis implies
$\langle0|w_{\boldsymbol v}\rangle=0$ and
$\langle w_{\boldsymbol v}|w_{\boldsymbol v}\rangle=|\boldsymbol v|^2$.
For $\boldsymbol v\ne0$, Eq.~\eqref{eq-si-qutrit-noise-quadratic-form} can therefore be written as
\begin{equation*}
 \mathcal D(\boldsymbol v)=|\boldsymbol v|^2P_{\boldsymbol v},
 \qquad
 P_{\boldsymbol v}=P_0+
 \frac{\ket{w_{\boldsymbol v}}\bra{w_{\boldsymbol v}}}{|\boldsymbol v|^2},
 \qquad P_{\boldsymbol v}^2=P_{\boldsymbol v}=P_{\boldsymbol v}^\dagger.
\end{equation*}
Here $P_{\boldsymbol v}$ is the orthogonal projector onto the two-dimensional span of
$\ket0$ and $\ket{w_{\boldsymbol v}}$.
Consequently, the eigenvalues of $\mathcal D(\boldsymbol v)$ are
$0,|\boldsymbol v|^2,|\boldsymbol v|^2$, and
\begin{equation*}
 0\preceq\mathcal D(\boldsymbol v)\preceq|\boldsymbol v|^2\mathbb I_3.
\end{equation*}
The same inequality holds trivially for $\boldsymbol v=0$.
The two positive terms in Eq.~\eqref{eq-si-qutrit-noise-quadratic-form} act on orthogonal subspaces, so the largest eigenvalue of their sum is $|\boldsymbol v|^2$.

\textit{Step 4: Bound for an arbitrary many-body state.}
Let $\rho_j=\Tr_{\ell\ne j}\rho_\Psi$ be the reduced density operator of site $j$.
It is positive with $\Tr\rho_j=1$, even when the probe is entangled.
Taking its expectation of the local operator inequality gives
\begin{equation*}
 0\le\Tr\!\left[\rho_j\mathcal D(\boldsymbol v)\right]\le|\boldsymbol v|^2.
\end{equation*}
Using the definition of $\mathcal Q$ in Eq.~\eqref{eq-output-covariance} and summing over sites then yields
\begin{equation*}
 0\le\boldsymbol v^T\mathcal Q\boldsymbol v
 =\sum_{j=1}^{N}\Tr\!\left[\rho_j\mathcal D(\boldsymbol v)\right]
 \le N|\boldsymbol v|^2.
\end{equation*}
Since this holds for every real $\boldsymbol v$ and $\mathcal Q$ is real symmetric, it is precisely the matrix inequality
\begin{equation}
 0\preceq\mathcal Q\preceq N\mathbb I_2.
 \label{eq-si-povm-noise-bound}
\end{equation}
Here $\preceq$ denotes positive-semidefinite matrix order.
No factorization of $\rho_\Psi$ is required: the intersite correlations are already included in $\mathcal C$, while $\mathcal Q$ depends only on the single-site reduced states.

For the product reference $|\Psi_{\rm p}\rangle=|0\rangle^{\otimes N}$, the local means vanish and
\begin{equation*}
 \langle0|Y_\alpha|0\rangle=0,\qquad
 \frac12\langle0|\{Y_\alpha,Y_\beta\}|0\rangle=\delta_{\alpha\beta},
 \qquad
 \langle0|\mathcal D_{\alpha\beta}|0\rangle=\delta_{\alpha\beta}.
\end{equation*}
Intersite covariances vanish on this product state, so
$\mathcal C_{\rm p}=N\mathbb I_2$ and $\mathcal Q_{\rm p}=N\mathbb I_2$.
Thus the upper bound in Eq.~\eqref{eq-si-povm-noise-bound} is attained, and
$\mathcal C^{\rm POVM}_{\rm p}=2N\mathbb I_2$.

\textit{Step 5: Conversion to phase-estimation covariance.}
Write $\widehat{\boldsymbol{\mathcal Y}}=
(\widehat{\mathcal Y}_x,\widehat{\mathcal Y}_y)^T$ and
$\boldsymbol\mu_0=\langle\boldsymbol{\mathcal Y}\rangle_{\boldsymbol\theta=0}$.
The first-moment identity implies the local response
\begin{equation*}
 \mathbb E_{\boldsymbol\theta}\!\left[\widehat{\boldsymbol{\mathcal Y}}\right]
 =\boldsymbol\mu_0+G\boldsymbol\theta+O(|\boldsymbol\theta|^2).
\end{equation*}
For nonsingular $G$, the calibrated linear estimator
\begin{equation*}
 \widehat{\boldsymbol\theta}
 =G^{-1}\left(\widehat{\boldsymbol{\mathcal Y}}-\boldsymbol\mu_0\right)
\end{equation*}
is locally unbiased: $\mathbb E_{\boldsymbol\theta}[\widehat{\boldsymbol\theta}]
=\boldsymbol\theta+O(|\boldsymbol\theta|^2)$.
At $\boldsymbol\theta=0$, the covariance of a linear transformation obeys
$\Cov(A\boldsymbol z)=A\Cov(\boldsymbol z)A^T$.
Applying this identity with $A=G^{-1}$ and
$\Cov(\widehat{\boldsymbol{\mathcal Y}})=\mathcal C+\mathcal Q$ gives, per preparation,
\begin{align}
 \Gamma_{\rm POVM}=G^{-1}(\mathcal C+\mathcal Q)(G^T)^{-1}=\Gamma_{\rm GW}+\Gamma_{\rm add},
 \quad \Gamma_{\rm add} =G^{-1}\mathcal Q(G^T)^{-1}.
 \label{eq-povm-sensitivity}
\end{align}
This is the covariance of the specified locally unbiased estimator.
The matrix bound also implies
\begin{equation*}
 0\preceq\Gamma_{\rm add}
 \preceq N G^{-1}(G^T)^{-1},
\end{equation*}
because multiplication on the left by $G^{-1}$ and on the right by its transpose preserves positive-semidefinite order.

\subsection{Finite-size results for the exact TAT echo}

\begin{figure}[H]
    \centering
    \includegraphics[width=0.5\linewidth]{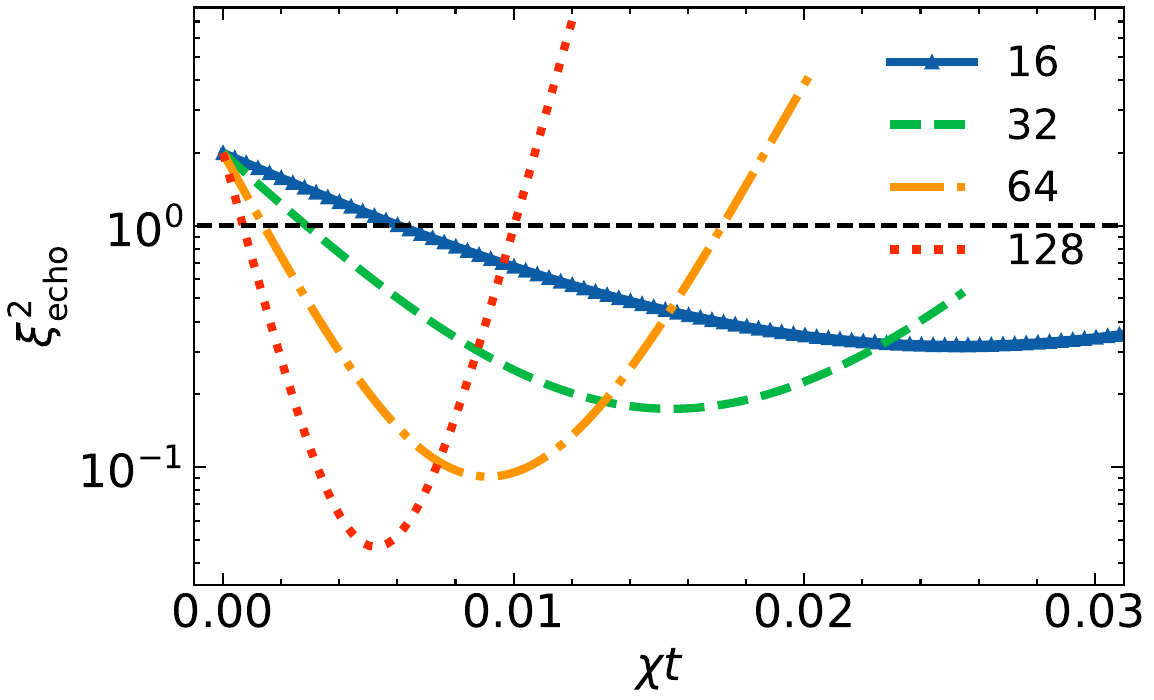}
    \caption{SQL-normalized cost $\xi_{\rm echo}^2=2N\Tr\Gamma_{\rm echo}$ for the exact collective TAT echo, followed by the fixed local POVM, versus interaction time $\chi t$ for $N=16,32,64,$ and $128$. The black dashed line at unity denotes the product-state SQL.}
    \label{fig-echo-readout}
\end{figure}

The finite-size results here use the collective TAT Hamiltonian and exact reversal from the main text. With $U_{\rm int}=e^{-iH_{\rm TAT}t}$, the sequence is $U_{\rm int}^\dagger U_{\boldsymbol\theta}U_{\rm int}\ket{\Psi_{\rm p}}$, followed by the fixed local POVM.
The quantities used in the calculation are
\begin{align}
 H_{\rm TAT}=\chi\sum_{\alpha=x,y} \{\mathcal X_\alpha,\mathcal Y_\alpha\},
 \quad 
 \mathcal X_\alpha(t) =e^{iH_{\rm TAT}t}\mathcal X_\alpha e^{-iH_{\rm TAT}t},
 \quad 
 G^{\rm echo}_{\beta\alpha} =i\langle[\mathcal X_\alpha(t),\mathcal Y_\beta] \rangle_{\Psi_{\rm p}},
 \label{eq-si-echo-slope}
\end{align}
with $\alpha,\beta\in\{x,y\}$.
At the operating point $\boldsymbol\theta=0$, the reversed evolution returns the state to $|\Psi_{\rm p}\rangle$, so the output covariance is the time-independent $\mathcal C^{\rm POVM}_{\rm p}=2N\mathbb I_2$ derived above.
Thus
\begin{equation}
 \Gamma_{\rm echo}
 =(G^{\rm echo})^{-1}\mathcal C^{\rm POVM}_{\rm p}
 [(G^{\rm echo})^T]^{-1}.
 \label{eq-si-echo-sensitivity}
\end{equation}
Using $\Gamma_{\rm SQL}=\mathbb I_2/(4N)$, we plot the SQL-normalized cost
\begin{equation}
 \xi_{\rm echo}^2
 =\frac{\Tr\Gamma_{\rm echo}}{\Tr\Gamma_{\rm SQL}}
 =2N\Tr\Gamma_{\rm echo}.
 \label{eq-si-echo-normalized-cost}
\end{equation}
At $\chi t=0$, $\xi_{\rm echo}^2=2$ because of the local POVM noise.
The interaction-enhanced slope subsequently drives the cost below unity for each size shown in Fig.~\ref{fig-echo-readout}.

\bibliography{ref}